\documentclass{aa}  

\usepackage{graphicx}
\usepackage[varg]{txfonts}
\usepackage{ulem}
\usepackage{natbib,twoopt}
\usepackage{color}
\usepackage{aecompl}
\usepackage{supertabular,booktabs}
\usepackage{float}
\usepackage{afterpage}
\usepackage{amssymb}
\usepackage{hyperref}
\hypersetup{
    colorlinks=true,
    linkcolor=blue,
    filecolor=magenta,      
    citecolor=cyan,
    anchorcolor=darkgreen, 
    urlcolor=cyan
    }

\usepackage{url} 
\usepackage{rotating}
\usepackage[flushleft]{threeparttable}
\usepackage{longtable}
\usepackage{subcaption}
\usepackage{amsmath}	
\usepackage{gensymb} 
\begin{document} 

\def\purple#1 {{\textcolor{purple}{#1}}\ }
\def\red#1 {\textcolor{red}{#1}}
\def\new#1 {{\bf #1 }}
\def\blue#1 {{\textcolor{blue}{#1}}\ }

\newcommand{\checkit}[1]{\textcolour{red}{#1} }
\newcommand{\noteit}[1]{\textcolour{magenta}{#1} }
\newcommand{\msun}{$M_\odot$}
\newcommand{\alpfe}{~[$\alpha$/Fe]~}
\newcommand{\teff}{$T_\mathrm{eff}$~}
\newcommand{\logg}{$\log(g)$~}
\newcommand{\feh}{[Fe/H]~}
\newcommand{\mh}{[M/H]~}
\newcommand{\gaia}{$Gaia$~}
\newcommand{\rgc}{$R_{\rm GC}$~}
\newcommand{\orcit}[1]{\protect\href{https://orcid.org/#1}{\protect\includegraphics[width=8pt]{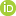}}}

\title{LAMOST meets \gaia: The Galactic Open Clusters }
\titlerunning{LAMOST OCs}
\authorrunning{X. Fu et al.}
    
\author{
Xiaoting Fu     \orcit{0000-0002-6506-1985}\inst{1,2,8},
Angela Bragaglia\orcit{0000-0002-0338-7883}\inst{2},
Chao Liu \inst{3},
Huawei Zhang    \orcit{0000-0002-7727-1699} \inst{4,1},
Yan Xu \inst {3},
Ke Wang     \orcit{0000-0002-7237-3856} \inst{1},
Zhi-Yu Zhang \inst{5,6},
Jing Zhong \inst{7},
Jiang Chang \inst{8},
Lu Li \inst{7,9,10},
Li Chen \inst{7,9},
Yang Chen \inst{11,3},
Fei Wang \inst {4,1},
Eda Gjergo \orcit{0000-0002-7440-1080} \inst {12},
Chun Wang \inst {13},
Nannan Yue \inst{1},
Xi Zhang \inst{7,9,14}
}

   
\institute{
        The Kavli Institute for Astronomy and Astrophysics at Peking University, 
        Yiheyuan Road 5, Beijing 100781, China \\ 
        \email{xiaoting.fu@pku.edu.cn}
    \and
        INAF - Osservatorio di Astrofisica e Fisica dello Spazio, 
        via Gobetti 93/3, 40129 Bologna, Italy  
    \and
        National Astronomical Observatories, Chinese Academy of Sciences,
        Datun Road A1, Beijing 100012, China  
    \and
        Department of Astronomy, Peking University, 
        Yiheyuan Road 5, Beijing 100781, China  
    \and
        School of Astronomy and Space Science, Nanjing University, 
        Nanjing 210093, China. 
    \and 
        Key Laboratory of Modern Astronomy and Astrophysics (Nanjing University), Ministry of Education,
        Nanjing 210093, People’s Republic of China 
    \and
        Key Laboratory for Research in Galaxies and Cosmology, Shanghai Astronomical Observatory, Chinese Academy of Sciences, 
        80 Nandan Road, Shanghai 200030, China. 
    \and
        Purple Mountain Observatory, Chinese Academy of Sciences, 
        Nanjing 210023, China 
    \and
       School of Astronomy and Space Science, University of Chinese Academy of Sciences, 
       No. 19A, Yuquan Road, Beijing 100049, China. 
    \and
       Centre for Astrophysics and Planetary Science, Racah Institute of Physics, The Hebrew University, 
       Jerusalem, 91904, Israel 
    \and
      Anhui University, 
      Hefei 230601, China. 
    \and
      School of Physics and Technology, Wuhan University, 
      Wuhan 430072, China 
    \and
        Tianjin Normal University,
        Tianjin 300387, China. 
    \and
      Changchun Observatory, National Astronomical Observatories, Chinese Academy of Sciences, 
      Changchun, China 
   }
   
\date{Received --, 2022; accepted --, 2022}


\abstract{Open Clusters are born and evolve along the Milky Way plane,
        on them is imprinted the history of the Galactic disc, including the chemical and dynamical evolution.  Chemical and dynamical properties of open clusters can
        be derived from photometric, spectroscopic, and astrometric data of
        their member stars.  Based on the photometric and astrometric data from
        the \gaia mission, the membership of stars in more than two thousand
        Galactic clusters has been identified in the literature.  The chemical
        (e.g. metallicity) and kinematical properties (e.g. radial velocity),
        however, are still poorly known for many of these clusters.  In synergy
        with the large spectroscopic survey LAMOST (data release 8) and \gaia
        (data release 2), we report a new comprehensive catalogue of 386 open
        clusters.  This catalogue has homogeneous parameter determinations of
        radial velocity, metallicity, and dynamical properties, such as orbit,
        eccentricity, angular momenta, total energy,
        and 3D Galactic velocity.  These parameters allow the first radial
        velocity determination and the first spectroscopic \feh determination
        for 44 and 137 clusters, respectively.  The metallicity distribution of
        majority clusters shows falling trends in the parameter space of the
        Galactocentric radius, the total energy, and the Z component of angular
        momentum -- except for two old groups that show flat tails in their own
        parameter planes.  
        Cluster populations of ages younger and older than
        500 Myrs distribute diversely on the disc. The latter has a spatial
         consistency with the Galactic disc flare.
        The 3-D spatial comparison between very young clusters (< 100 Myr) and nearby molecular clouds
        revealed a wide range of metallicity distribution along the Radcliffe
        gas cloud wave, indicating a possible inhomogeneous mixing or fast star
        formation along the wave. This catalogue would serve the community as a
        useful tool to trace the chemical and dynamical evolution of the Milky Way.  }
  
\maketitle

   
\section{Introduction}

The \gaia mission  \citep[see][]{GaiaCollaboration2018a, gedr32021, dr3} is
revolutionising our knowledge of the Milky Way (MW) with its very precise and
accurate astrometry and photometry of more than 1.8 billion stars. Many of
these stars are found in stellar clusters, which are important components of
the Galaxy. In particular, open clusters (OCs) could trace the formation
history and chemical properties of the Galactic disc. They also provide very
useful tests of stellar evolution models \citep[see e.g.][for two examples
based on results obtained by the $Gaia$-ESO Survey]{Semenova2020,Magrini2021}.
Characterizing OCs is therefore a fundamental task. Such a process includes discovering
them, separating cluster populations from underlying field interlopers,
measuring radial velocities (RV) and chemical abundances, and deriving
distances and ages, etc.  All these tasks can be done more effectively by
combining the \gaia results with ground-based data \citep[see
e.g.][]{Bragaglia2018a, Carrera2019, Zhong2020, Casali2020a,
AlonsoSantiago2021, Spina2021}.

The most commonly used catalogues of OC properties before \gaia are
\citet[][and its web updates]{Dias2002} and \cite{Kharchenko2013}, in which
 2000 to 3000 objects are considered, respectively. 
In the \gaia era, 
mostly thanks to the precise astrometric information (parallax, $\varpi$' and
proper motion, PM) and \gaia's full-sky coverage. For instance, membership of OC
stars has been studied using \gaia data release 1 (DR1) and the Tycho-Gaia Astrometric Solution (TGAS) by \citet{GaiaCollaboration2017,CantatTGAS,Yen2018,Randich2018}.
Using \gaia DR2, member stars in known OCs are identified
\citep[e.g.][]{CantatDR2, Cantat-Gaudin2020a, Jackson2022} and new OCs have
been discovered \citep[e.g.][]{CantatPerseus, CastroGinard2018,
CastroGinard2019, CastroGinard2020, Beccari2018, Ferreira2019, liupang2019}.
The Early Data Release 3 of \gaia, has already been used to detect new OCs
\citep[e.g.][]{CastroGinard2022}. In particular, \citet{Cantat-Gaudin2020b}
combine highly reliable cluster membership in OCs -- known and identified by
other works -- and estimate the age, distance, and reddening for $\sim$ 2000
OCs; their data set will be used in the present paper.  All these works,
together with the revision of the OC census, also mean that many candidate
clusters have not been confirmed, see for instance the discussions in
\citet{Kos2018} and \citet{CantatDR2}.

The kinematical information of OCs is based on radial velocity measurements of
stellar spectra.  While \gaia spectroscopic capabilities are limited \citep[see
e.g.][]{Sartoretti2018,Katz2019}, its instrument Radial Velocity Spectrometer
\citep[RVS, ][]{rvs} collected data for several million bright stars. Matching the \gaia RVs to
OC members (derived by \citealt{CantatDR2}), \citet{Soubiran2018} could
determine average RVs for nearly 900 clusters and derive their kinematics. The
work has been extended by \citet{Tarricq2021}, who also included data from
ground-based surveys.
 
\gaia data have already been extensively adopted to clean colour-magnitude
diagrams (CMD) of OCs, and to derive more precise ages (and distance, if not
computed directly from the cluster $\varpi$), see for instance
\citet{Randich2018,Yalyalieva2018,Dias2018,Choi2018} and the method described
in \citet{li2022}. An extensive derivation of cluster ages can be found in
\citet{GaiaCollaboration2018b,Bossini2019}. In the latter paper, a Bayesian
code was applied to the list of clusters in \citet{CantatDR2} and they were
able to obtain excellent age results  for about 270 OCs. However, a common limitation of these works is the
absence of metallicity information in the majority of Galactic clusters, which
introduced degeneracies with reddening and age.

In fact, metallicity measured with high-resolution spectroscopy is available
only for about 10\% of the whole OC population.
This low percentage of highly resolved OC metallicity is not due to a shortage of studies. 
To cite only a few, \citet{Magrini2018a} combined \gaia-ESO Survey data with compilations from \citet{Netopil2016}, 
while \citet{Donati2015,Reddy2016,Reddy2019,Casamiquela2017,Smiljanic2018a, Bragaglia2018b, Casali2020b} 
are based on private projects such as BOCCE, OCCASO, and SPA. 
Although \gaia will obtain the metallicity and some
elemental abundances on a grand scale with RVS, to derive more detailed
elemental abundances, ground-based surveys are
fundamental especially in faint clusters. Examples are the high-resolution \gaia-ESO
\citep{Gilmore2012,Randich2022}, APOGEE \citep{Majewski2017}, GALAH
\citep{DeSilva2015}, and the low-resolution LAMOST \citep[Large Sky Area
Multi-Object Fiber Spectroscopic Telescope,][]{Cui2012, Deng2012, Zhao2012}
surveys. Future surveys such as WEAVE \citep{Dalton2012} and 4MOST
\citep{deJong2019} are also planning to observe the OC population.
 
The \gaia-ESO Survey targeted on purpose 62 OCs, observing a few hundred to
thousands of stars at an intermediate resolution, and roughly tens of members at high resolution in each of them \citep{Bragaglia2021, Randich2022}. 
About 20 more clusters from the ESO archive observations were
also re-analysed homogeneously and included in the data release \citep[see for
instance][]{Magrini2017, Bragaglia2021,Randich2022}. The earlier data release
of the main GALAH survey does not have OCs \citep[see e.g.][]{Buder2018}, while
their latest release covers 75 OCs \citep[DR3,][]{Buder2021} which were also analysed
together with the APOGEE OCs to provide a homogeneous set \citep{Spina2021}.
The APOGEE OC samples are mainly presented within the OCCAM program \citep[Open
Clusters Chemical Abundance and Mapping,][]{Donor2018, Donor2020}, with a few
to a few tens of stars in each OC.  However, more OC stars have been
serendipitously observed both by GALAH and by APOGEE, as found by \citet{Carrera2019}. They
cross-matched the OC member stars as defined by \citet{CantatDR2} with the
survey data releases and were able to retrieve RVs, metallicities, and chemical
abundances for more than 100 OCs, many of them without previous determinations.

The same technique can be applied to the LAMOST survey, to extend the number of
OCs with measured RV and metallicity, and investigate their chemical,
kinematical, and dynamical properties on galactic scales, with a catalogue of
homogeneous analysis. Based on a previous data release (LAMOST DR5) and an
earlier \gaia OC membership catalogue \citep{CantatDR2}, \citet{Zhong2020}
explored properties of 295 clusters and discussed their metallicity
distributions in the MW.  The latest LAMOST data release, DR8{\footnote{\url{http://www.lamost.org/dr8/}}}, includes 10,388,423 stellar spectra in total,
which were observed between October 24th, 2011 and May 27th, 2020. This work is
an updated and extended version of \citet{Zhong2020} on the LAMOST OC
investigations based on \gaia.

This paper is organised as follows. 
In section \ref{sec:data} we introduce the LAMOST and \gaia data adopted in this work, together with quality control methods.
The catalogue results after the quality control, i.e. the radial velocity and metallicity of the LAMOST OCs, are described in Section \ref{sec:res}. 
In Section \ref{sec:dis} we discuss the Galactic metallicity distribution obtained with our LAMOST OC catalogue, the dynamical properties of OCs,
and the connection with the Galactic molecular clouds.
Lastly, the main conclusions of this paper are summarized in Section \ref{sec:sum}.

\section{Data and quality control}
\label{sec:data}

The Large Sky Area Multi-Object Fiber Spectroscopic Telescope (LAMOST, Guo Shou
Jing Telescope), located in Xinglong, China, is a quasi-meridian reflecting
Schmidt telescope with an effective aperture of $\sim$4 m. With its 4,000
fibres, it is one of the most efficient spectroscopic telescopes. In the low
resolution mode (R=1800), its limiting magnitude is $r = 19$ mag. 

In this work we adopt LAMOST DR8 low-resolution catalogue with stellar
parameters (i.e. LRS A, F, G, and K Type Star Catalog).  In total, 6,478,063
spectra are published in the original LAMOST catalogue, including 100,468
A-type, 1,983,821 F-type, 3,249,746 G-type, and 1,144,028 K-type stars.  All
spectra in this catalogue have a criterion of $g$ band signal-to-noise ratio,
S/N$>6$ for dark night observations, or S/N$>$15 for bright night observations.
RV and stellar parameters (i.e. effective temperature \teff, surface gravity
\logg, and iron abundance \feh) in this catalogue are determined with the
official LAMOST Stellar Parameter pipeline \citep[LASP,][]{lasp}, which uses
ATLAS9 atmosphere models \citep{atlas9} and the \citet{Grevesse1998} Solar
abundances.

\begin{figure}
    \centering
    \includegraphics[width=.48\textwidth]{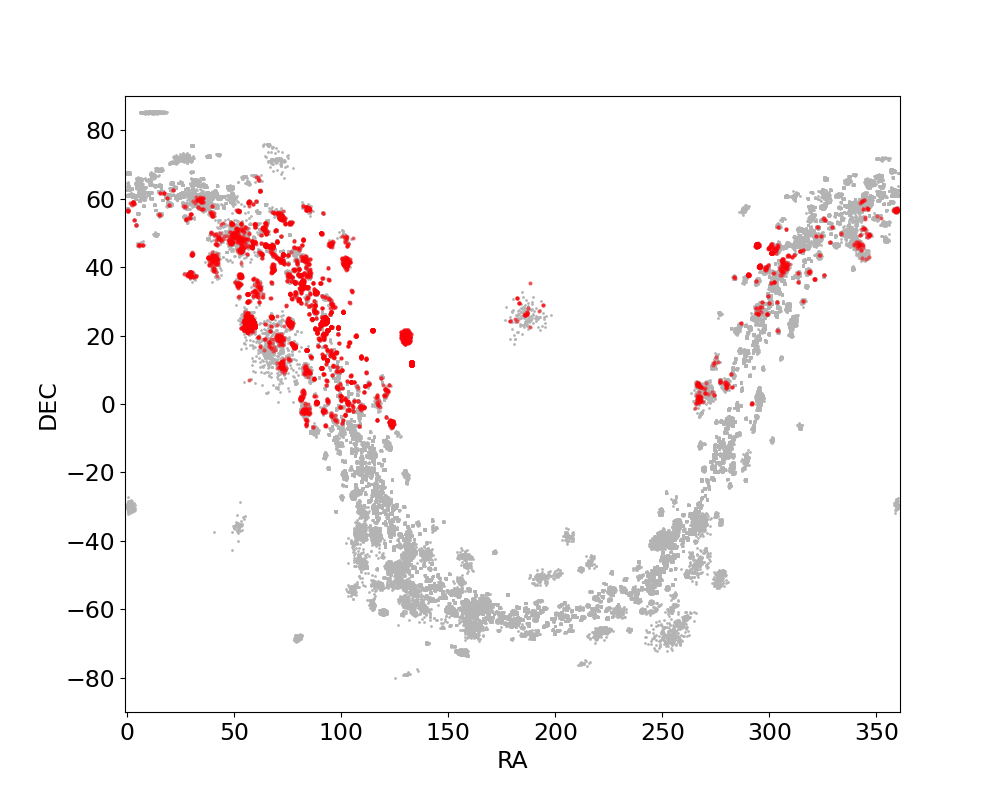}
    \caption{The sky coverage of the LAMOST open cluster (OC) members (red dots) and all the \citet{Cantat-Gaudin2020b} OC member stars (grey dots). 
    }
    \label{fig:radec}
\end{figure}

To obtain RV and stellar parameters of OC member stars, we cross-matched the
LAMOST and \citet{Cantat-Gaudin2020b} catalogues, keeping only members with
probabilities $>$ 70\%. For each star, we matched the above two catalogues with
its \gaia \texttt{source\_id}  .  The \citet{Cantat-Gaudin2020b} catalogue has
already the \gaia DR2 \texttt{source\_id}, so the first step of our procedure
was to match each LAMOST star with the \gaia DR2 data.  We used the CDS X-match
service in TOPCAT \citep{topcat, topcat2017} to consider both the PM and the
epoch of \gaia stars.  Targets are identified with their R.A. and Dec.
coordinates within 3.5 arcsec, because the fibre size of LAMOST is 3.3 arcsec
\citep{Zhao2012} and the dome seeing of LAMOST is sometimes slightly larger
than the fibre scale \citep{Luo2015}. 
In the X-match procedure, all \gaia sources in the matching radius are considered because light from different sources cannot be resolved within the LAMOST fibre size.
In some cases more than one \gaia sources are matched for a LAMOST observation.
However, our determinations of the clusters' velocity and \feh are not affected 
because a Monte Carlo sampling selection will be applied in the procedure (see details in Sec. \ref{sec:rv}).
After getting the \gaia DR2 \texttt{source\_id} of each star in the LAMOST catalogue, 
we used the \texttt{source\_id} as the identification to cross-match the LAMOST-Gaia table
with the \gaia OC member star catalogue \citep{Cantat-Gaudin2020b}.  In total,
7,570 stars from 386 clusters are in common. The spectra of these stars have
high S/N: the mean S/N in $g$ band is 109 and $\sim$ 78\% stars have $g$ band
S/N $>$30. Fig. \ref{fig:radec} shows the sky coverage of the LAMOST OC members
(red dots). For comparison, all OC member stars of \citet{Cantat-Gaudin2020b}
are overlaid as grey dots. Since LAMOST operates from the northern hemisphere,
the southernmost matched stars have a declination of $\sim-7 ^{\circ}$.  Among
all the LAMOST OCs, 308 OCs have at least two matched stars, the other 78
clusters have only one star in the matched table.
We keep all these clusters for further analysis. In principle, even one star can represent the cluster because OCs are very homogeneous in their chemistry and radial velocity.

To provide accurate chemical and kinematical information for these selected
LAMOST OCs, we perform quality control on radial velocity and stellar
parameters, respectively.

\subsection{Radial velocity quality control}
\label{sec:rv}

The greatest advantage of LAMOST is the vast number of spectra, but the low
resolution makes the stellar RV uncertainty not negligible compared to studies
with high-resolution spectra. The median value of the RV uncertainty for our
matched LAMOST OC member stars is 6.36$\pm$3.20 km~s$^{-1}$, which is comparable
to or even larger than the typical RV dispersion of a cluster.  For instance,
the RV dispersion of clusters NGC\,2516, NGC\,6705, and NGC\,6633 are 1.0, 1.6, and 1.5 km~s$^{-1}$, respectively, as reported
in the \gaia-ESO survey based on high resolution spectra \citep[see
e.g.][]{Magrini2017},.

To derive the average RV of clusters, the RV dispersion of each cluster, and a
proper RV member quality control, we apply a Monte Carlo (MC) method, by
considering each individual RV and its corresponding uncertainty from the
LAMOST DR8 catalogue.  For simplicity we assume that all RV uncertainties
follow Gaussian distributions and then we randomly sample RV of every member star
in each cluster 5,000 times.  This allows us to derive the mean and median
of all the sampled RV values.  Member star with RV measurements within 2
$\sigma$ of the MC RV sampling are marked with a quality control flag of {\sc
flag = 1}, while those with $>$ 2 $\sigma$ are marked as {\sc flag = 0}.  The
mark of {\sc flag = 1} means we use the RV values of the star to determine the
cluster radial velocity.

\begin{figure}
    \centering
    \includegraphics[width=.48\textwidth]{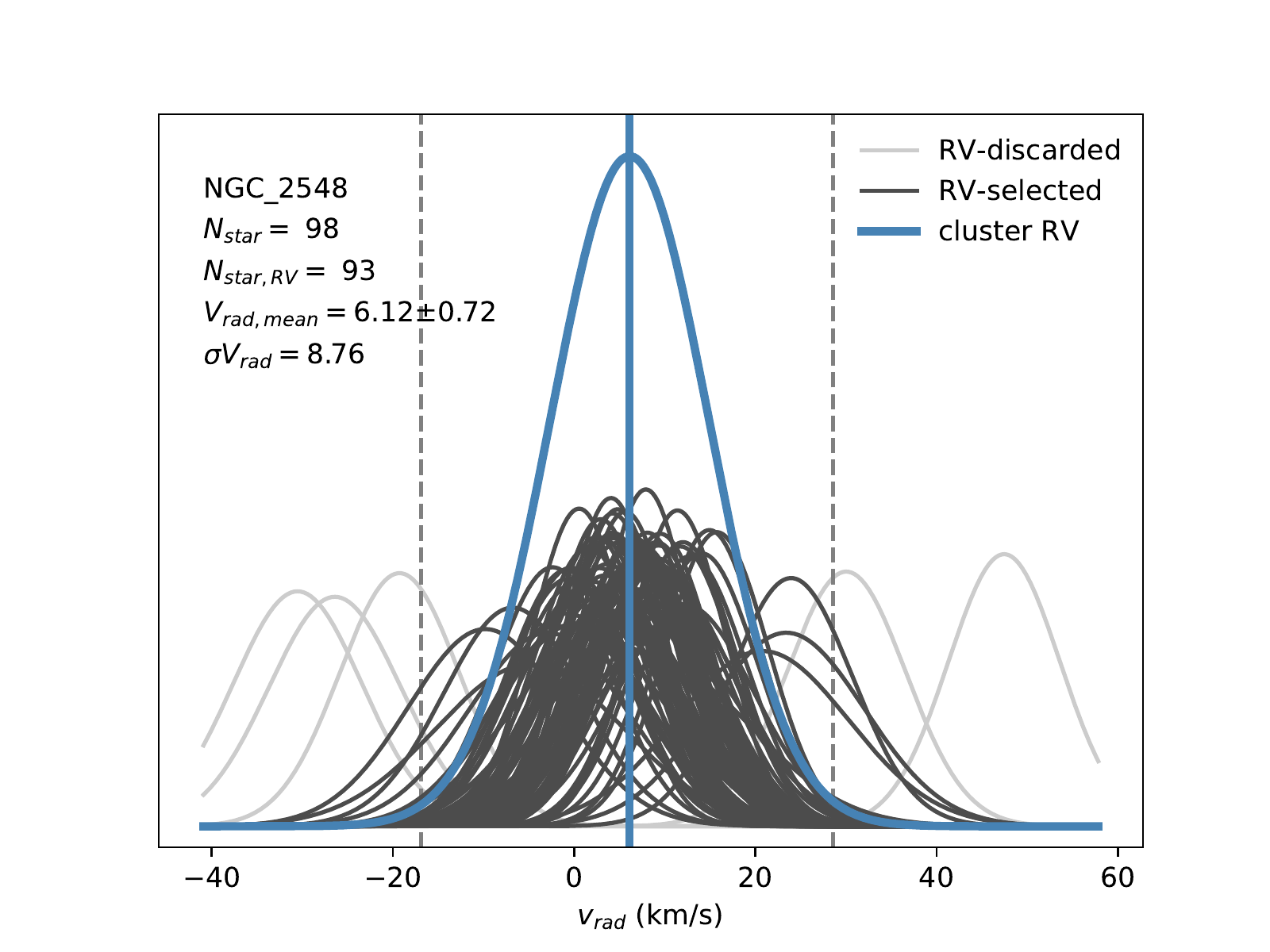}
    \caption{NGC\,2548 as an example of the RV selection.
    The RVs of all the crossed-matched member stars in the cluster are plotted assuming a normal distribution.
    The light grey colour curves denote stars we discard for further analysis ({\sc flag = 0}), while the dark grey ones are the selected stars. 
    The two vertical dashed lines mark the 2-$\sigma$ selection criterion.
    The Monte Carlo model result of all the selected stars is illustrated with the blue Gaussian curve.
    Their mean RV value ($V_{\rm rad,mean}$), 
    the 1-$\sigma$ RV dispersion ($\sigma V_{\rm rad}$), as well as  the number of member stars before ($N_{\rm star}$) and after the RV selection ($N_{\rm star,RV}$), are shown in the legend. 
    }
    \label{fig:rv}
\end{figure}

Fig.~\ref{fig:rv} shows an example of the RV quality control in cluster
NGC~2548, where RV of all matched stars are plotted as Gaussian profiles.
Dark-grey curves show selected RV member ({\sc flag =1}), while light-grey
curves are stars we discard ({\sc flag=0}) for the cluster radial velocity
determination. The thick blue curve is the Gaussian fit of all the stars with
{\sc flag=1} in this cluster.  The thick blue vertical line represents the mean
radial velocity of the cluster ($V_{\rm rad, mean}$).  The two vertical dashed
lines mark the 2-$\sigma$ departure from $V_{\rm rad, mean}$, which are adopted
as thresholds of {\sc flag=1} star selections. In the top left region of the
Fig. \ref{fig:rv}, we show numbers of total matched stars in the cluster
($N_{\rm star}$), numbers of RV-selected ({\sc flag=1}) stars ($N_{\rm star,
RV}$), the cluster mean radial velocity ($V_{\rm rad, mean}$), and the
corresponding standard deviation ($\sigma\, V_{\rm rad}$), which is adopted as
the cluster radial velocity uncertainty, respectively. 
In most clusters, $V_{\rm rad, mean}$ and the median radial velocity of the cluster $V_{\rm rad, med}$ are very similar to each other, with a mean absolute
difference of $\sim$ 0.85 km~s$^{-1}$. Indeed, 76\% of clusters have a
difference smaller than this mean value. In the following part of this paper,
we use $V_{\rm rad, med}$ to discuss the property of cluster radial velocity $V_{\rm rad}$.

\begin{figure}[!htb]
    \centering
    \includegraphics[width=.48\textwidth]{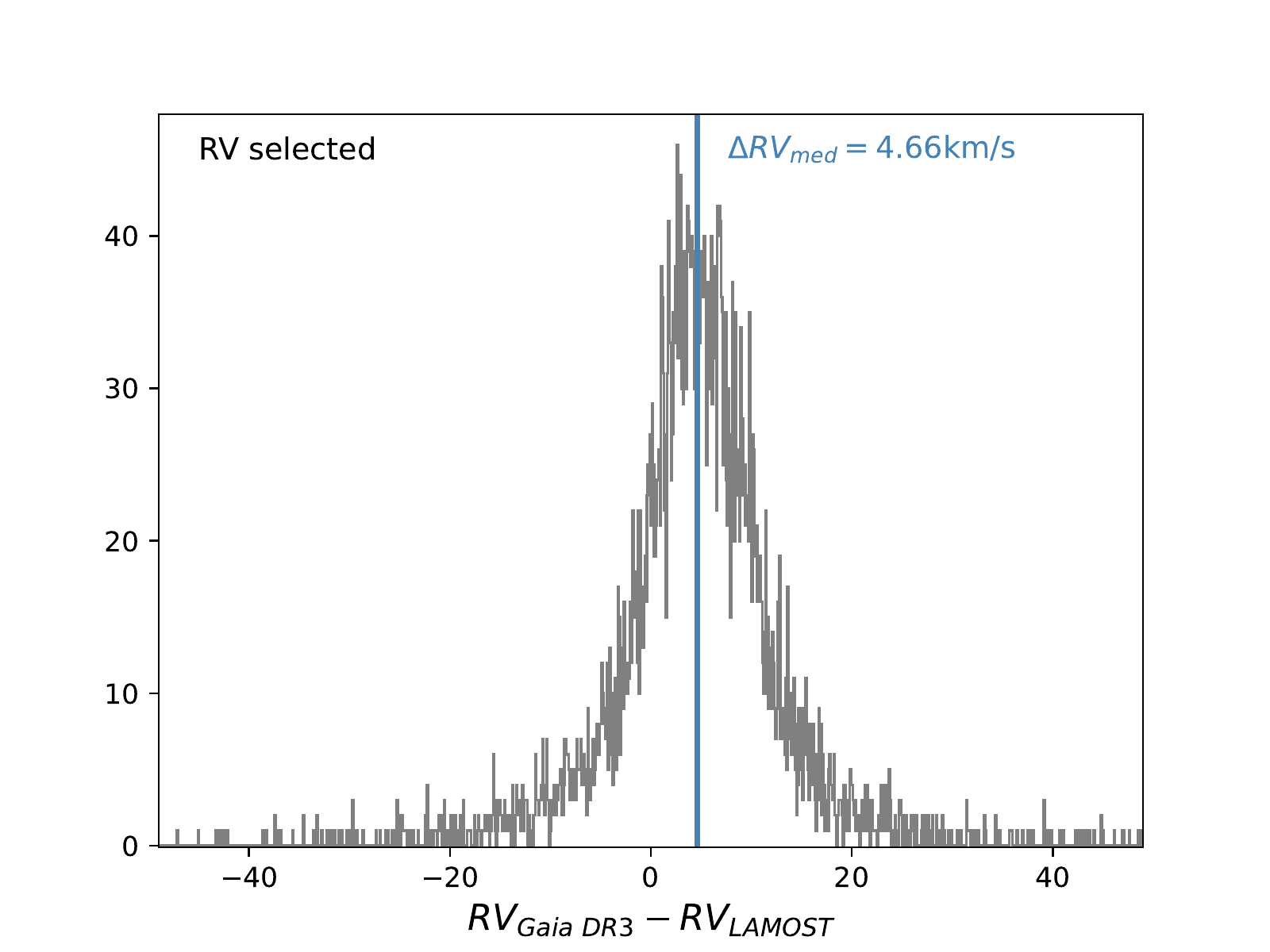}
    \caption{Histogram of the radial velocity difference between \gaia DR3 measurements and LAMOST measurements for all the RV selected members.
    The median value of the difference $\Delta RV$ = RV$_{Gaia~DR3}$ - RV$_{LAMOST}$ is 4.66 km~s$^{-1}$.
    }
    \label{fig:drv}
\end{figure}

Stellar RV measurements of LAMOST are known to have a systematic offset compared to higher resolution data, with a value
\footnote{
For instance, the offset value is 
3.78 km~s$^{-1}$ compared to SIMBAD literature results \citep[][LAMOST DR1]{Gao2015},
4.54 km~s$^{-1}$ compared to APOGEE DR14  \citep[][LAMOST DR3]{Anguiano2018},
 4.9 km~s$^{-1}$ compared to GALAH DR2 \citep[][LAMOST DR5]{Zhong2020},
and  4.4 km~s$^{-1}$ compared to RAVE DR3 \citep[][LAMOST LSP3 based on DR1 spectra]{Xiang2015}.   }
about 4-5 km~s$^{-1}$.
In the ``Survey of Surveys'' work by \citet{Tsantaki2022}, where they compare
RV measurements from different survey data, the median and mean difference
between \gaia DR2 and LAMOST DR5 (RV$_{Gaia~DR2}$-R$_{LAMOST}$) are 4.97 and 5.18
km~s$^{-1}$, respectively. In Fig. \ref{fig:drv} we show the RV measurement
difference ($\Delta$RV = RV$_{Gaia~DR3}$-R$_{LAMOST}$) distribution of all the
{\sc flag=1} OC members.  The $\Delta$RV have a median value of 4.66
km~s$^{-1}$, mean value of 4.82 km~s$^{-1}$, and a standard deviation of 12.29
km~s$^{-1}$. All these values based on cluster member stars after the RV
quality control are similar to the raw ones reported by \citet{Tsantaki2022}
before their RV correction.

\subsection{Stellar parameter quality control}
\label{subsec:ms}

Similar to the quality control of $V_{\rm rad}$ described in Sec. \ref{sec:rv},
it is also necessary to control the qualities of stellar parameters before
further analysis.  This control procedure is based on not only finding  \feh
outliers, but also checking member star evolution.  In principle, the surface
gravity \logg and effective temperature \teff of member stars in the same
cluster should follow the evolution of a simple stellar population in the Kiel
diagram, where the stellar \logg value is an index of the evolutionary phase.
The \feh of these stars, on the other hand, should be almost a constant in
different evolutionary phases \footnote{This is not strictly true, as \feh
has been shown to vary due to atomic diffusion. However, the variations are
within about 0.1 dex, see for instance \citet{BertelliMotta2018, Semenova2020}
for two well studied OCs.}.

\begin{figure}
    \centering
    \includegraphics[width=.48\textwidth]{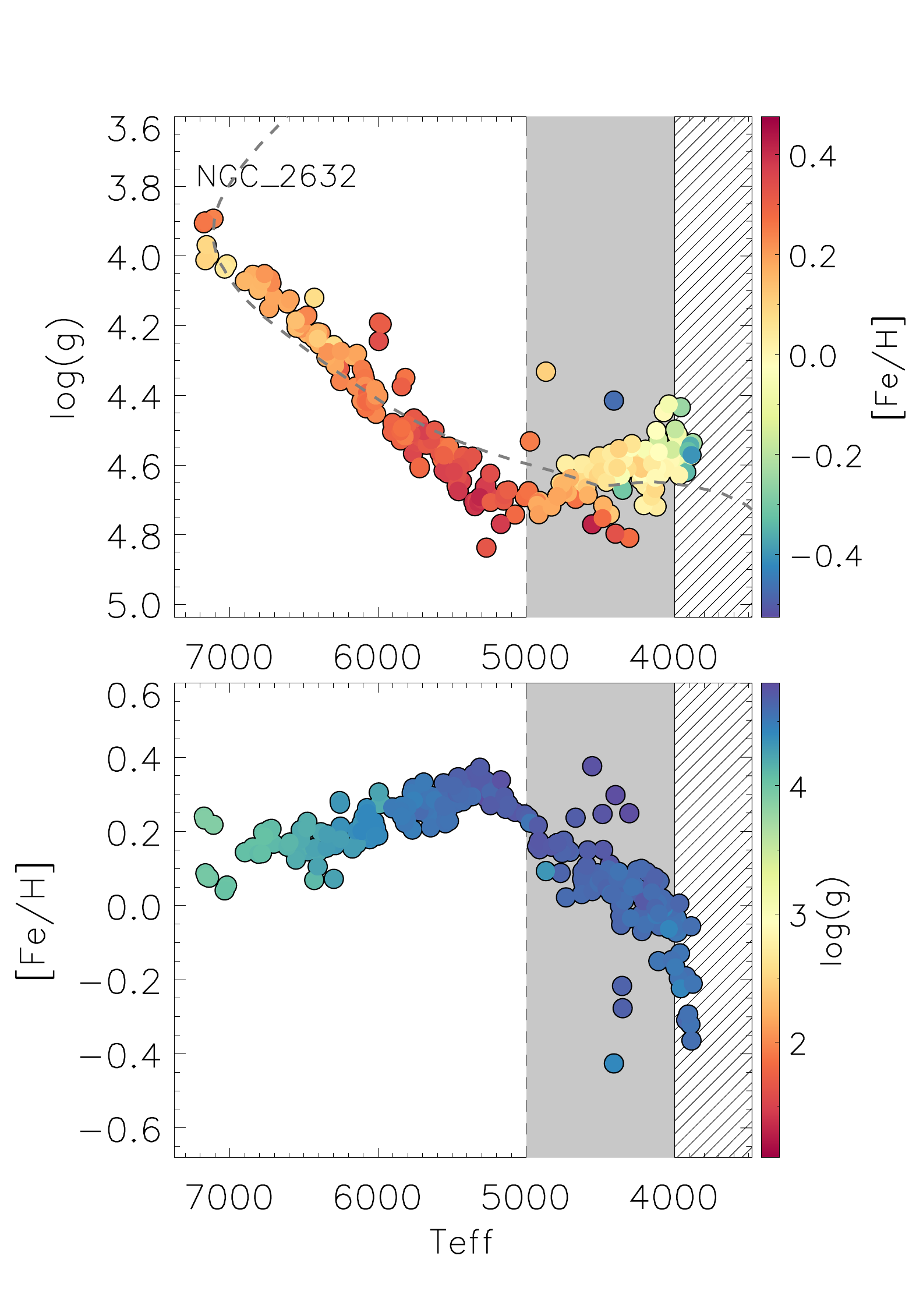}
    \caption{Kiel diagram (upper panel, colour-coded with \feh) and 
    \feh as a function of \teff (lower panel, colour-coded with \logg)
    of cluster NGC\,2632.
    For reference, a \texttt{PARSEC} theoretical isochrone of Z=0.013, log(age)=9.3 
    is plotted in the Kiel diagram.
    The line-shaded region marks \teff< 4000\,K,
    and the grey part are 5000\,K<\teff<4000\,K.
    }
    \label{fig:prob}
\end{figure}

Therefore, such quality control could serve as  useful examinations, or even
calibrations, to check the accuracy of stellar parameter analysis.  In these
quality control process to our LAMOST OC member stars, we find a systematic
issue for the  \logg and \feh parameters of cool main sequence stars.
Fig.~\ref{fig:prob} presents cluster NGC\,2632 as an example of such an issue.
We find that main sequence stars with \teff $\lessapprox$ 5000\,K do not follow the
main sequence evolution trend. 
For reference, a theoretical isochrone from \texttt{PARSEC~v1.2S}  
\footnote{\url{http://stev.oapd.inaf.it/cgi-bin/cmd_3.6}}
\citep{Bressan2012, chen2014, chen2015, tang2014} with a metallicity of Z=0.013 and log(age)=9.3 is plotted in the Kiel diagram (the upper panel of Fig.\ref{fig:prob}).
Most of these stars have a \logg value lower than the
expected values (see stars in the grey part of the upper panel of Fig. \ref{fig:prob}).
Their \feh values are also abnormally lower than that of other member stars (see the lower panel of Fig.~\ref{fig:prob}).

On the other hand, giant stars in this temperature regime seem not affected.
Member stars' \feh values are almost a constant for all giants as well as dwarfs at
\teff$\gtrsim$5000\,K.  Only the dwarf stars show a decline tend at lower temperature,
which is seen in all of our clusters that have dwarf stars in this temperature
range.  To select OC member stars with a more secure \feh determination, we
discard dwarf stars cooler than 5000\,K in our cluster \feh calculation.

It is always difficult to obtain stellar parameters for very hot and very cool
stars.
There are few metal lines that can be used in hot star spectra, and the continuum is relatively difficult to derive.
For stars with cool temperature, molecular lines dominate their spectra and make the  analysis challenging \citep[see discussions in ][]{jofre2019}. 
Therefore, the second step of our stellar parameter quality control process is excluding \feh of member stars with \teff $>$7500\,K and \teff$<$4000\,K for the cluster metallicity determinations.
The hot star criterion \teff=7500\,K is adopted from LAMOST hot star stellar parameter work \citep{Xiang2021}.

After the two-step quality control process described above, we end up with 355
OCs for further cluster \feh determination.  The selected member stars are
marked with {\sc flag=12} in our output catalogue, which means the parameters
of these stars are good for both RV and \feh determination of the cluster.
Among these clusters, 203 ones have at least three members with {\sc flag=12}. 

We then calculate the cluster \feh together with the corresponding uncertainty
and scatter using a Monte Carlo method. 
The mean stellar \feh uncertainty of our OC member stars from the pipeline LASP is 0.07 dex.
Assuming \feh uncertainty of member
stars follow Gaussian distributions, we apply a random \feh sampling of 5,000
times to each cluster, and obtain the median \feh$_{med}$, mean \feh$_{mean}$,
and standard deviation $\sigma$\feh of all the sample values.  The absolute
difference between the cluster median \feh$_{med}$ and mean \feh$_{mean}$
values are very small, with a mean difference of $\sim$0.01 dex. About 70\% of
clusters have difference smaller than this value.  In the rest of this
paper, we adopt the median value \feh$_{med}$ to discuss the cluster \feh, and
take the sample standard deviation $\sigma$\feh as the cluster \feh
uncertainty. 

\begin{figure}
    \centering
    \begin{subfigure}[]{0.48\textwidth}
        \includegraphics[width=\textwidth]{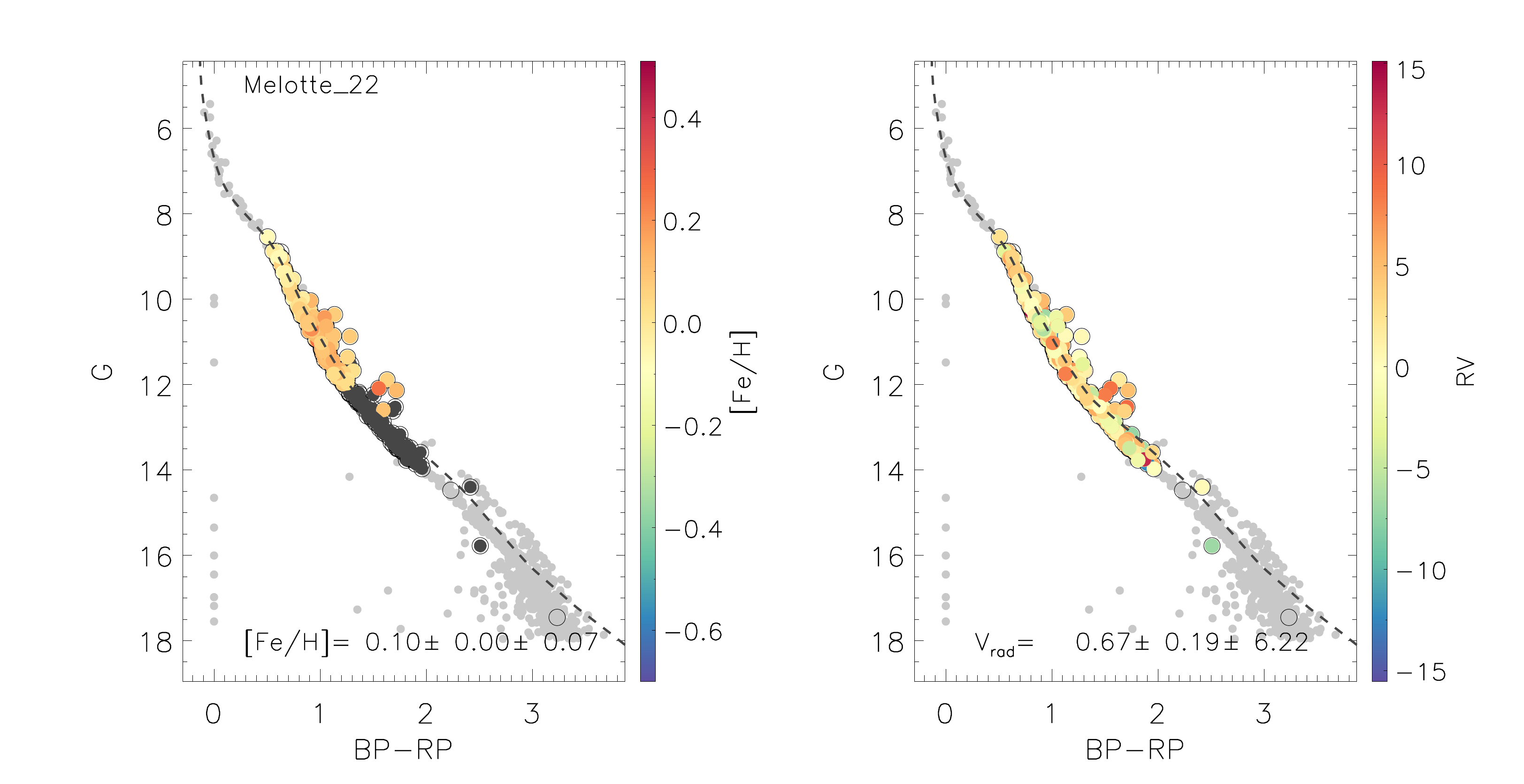}
        \label{fig:ngc2682}
    \end{subfigure}
    \begin{subfigure}[]{0.48\textwidth}
        \includegraphics[width=\textwidth]{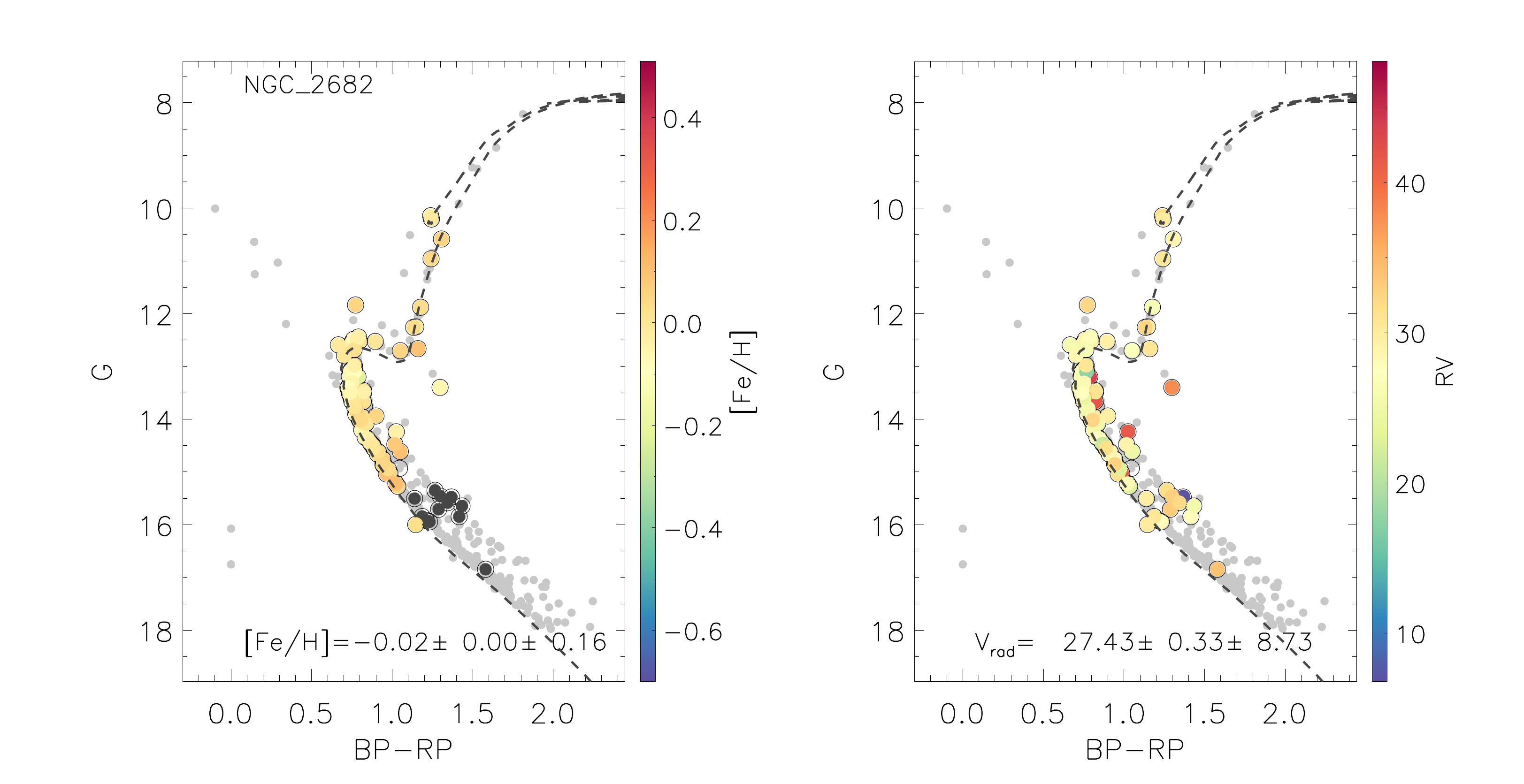}
        \label{fig:n752}
    \end{subfigure}
    \caption{Colour-magnitude diagrams of Melotte\,22 and NGC\,2682 as examples
            of the clusters with \feh in the literature.  The grey filled
            circles are high quality member stars from
            \citet{Cantat-Gaudin2020b}, and open circles are all matched stars
            with LAMOST RV measurements.  In the right panel stars good for
            the cluster $V_{\rm rad}$ determination  ({\sc flag=1}) are
            colour-coded with their RV values.  In the left panel, the colour-filled dots are stars good for \feh determination ({\sc flag=12}), and the black ones are stars without \feh determination or with discarded \feh.
            The dashed curve in each sub-figure is the \texttt{PARSEC} isochrone with the \citet{Cantat-Gaudin2020b} cluster parameters. 
    }
    \label{fig:cmd}
\end{figure}

Fig.~\ref{fig:cmd} shows the CMD of  member stars with \gaia DR2 photometry of
two very typical OCs in our catalogue. They are Melotte 22 (Pleiades) in the
upper two panels and NGC\,2682 (M67) in the lower two panels.  The left panel
of each cluster is colour-coded with member star \feh, which means all the
coloured ones are {\sc flag=12} stars.  The right panel of each cluster is
colour-coded with member star RV, marking all the {\sc flag=1} stars in the
cluster.  The derived cluster $V_{\rm rad}$ and \feh values based on the {\sc
flag=1} and {\sc flag=12} members, respectively, are shown in the figure.

\section{Results}
\label{sec:res}

\subsection{Radial Velocity}
\subsubsection{Clusters with  V${\sc _{rad}}$ in the literature}
\label{subsec: knownrv}

Catalogues of the Galactic OC radial velocity have been compiled in many works
\citep[see e.g. ][]{Dias2002,Kharchenko2013, Conrad2014, Soubiran2018,
Dias2021, Tarricq2021}.  We compare our cluster  V${\sc _{rad}}$ results to the two
most recent and most complete large catalogues, from \citet{Tarricq2021} and
\citet{Dias2021}, together with the previous LAMOST OC catalogue from
\citet{Zhong2020}.

\begin{figure*}[!htb]
    \centering
    \begin{subfigure}[]{0.32\textwidth}
        \includegraphics[width=\textwidth]{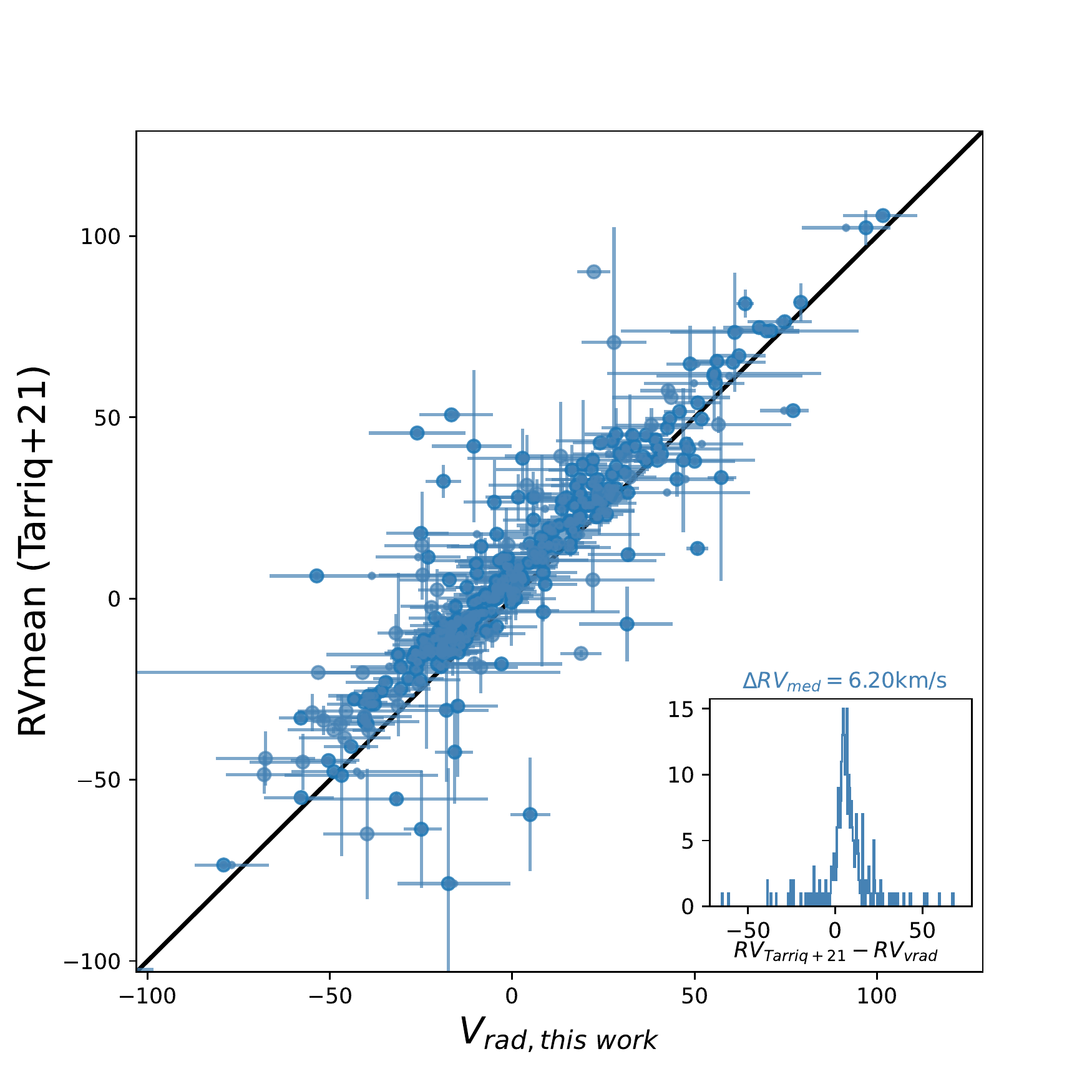}
    \end{subfigure}
    \begin{subfigure}[]{0.32\textwidth}
        \includegraphics[width=\textwidth]{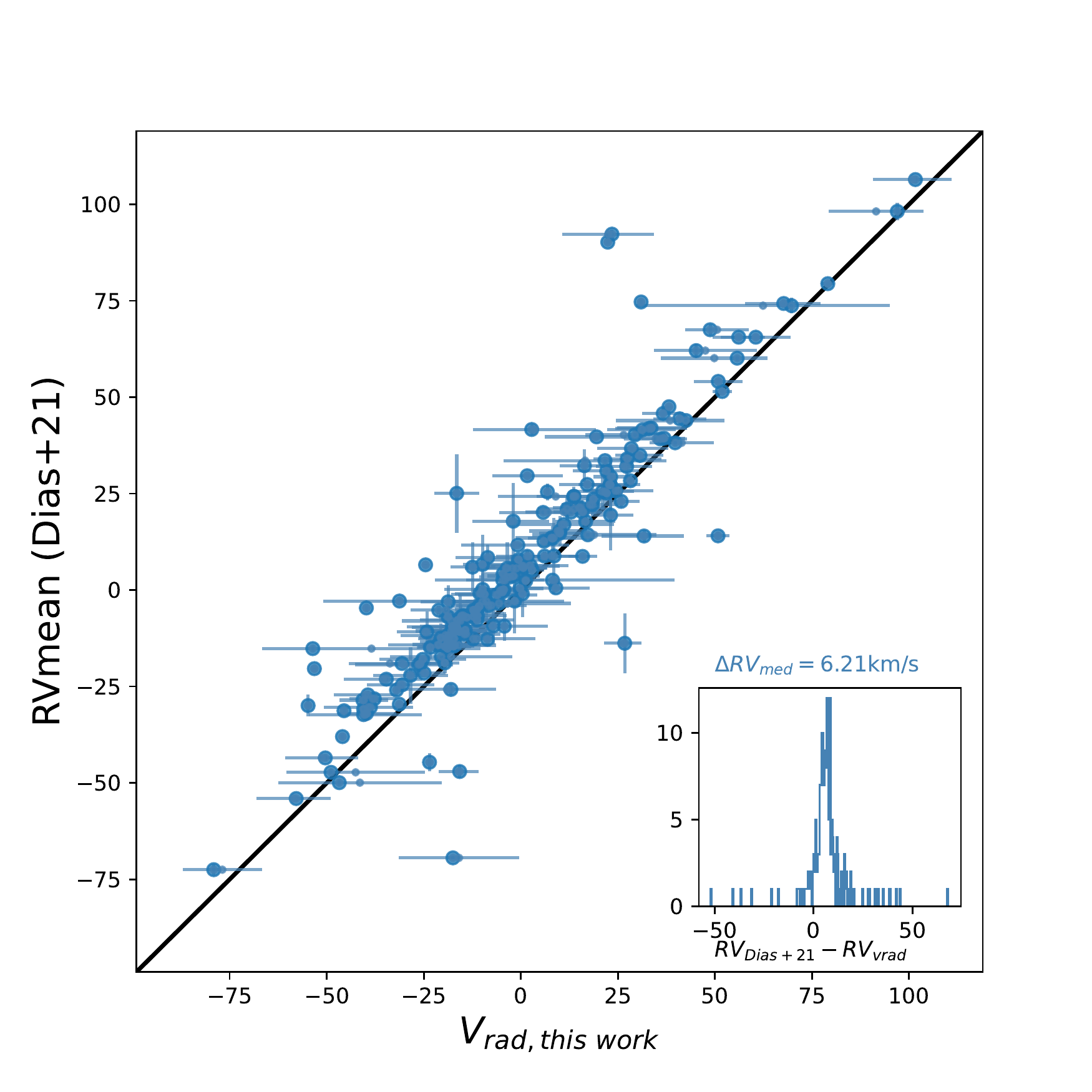}
    \end{subfigure}
    \begin{subfigure}[]{0.32\textwidth}
        \includegraphics[width=\textwidth]{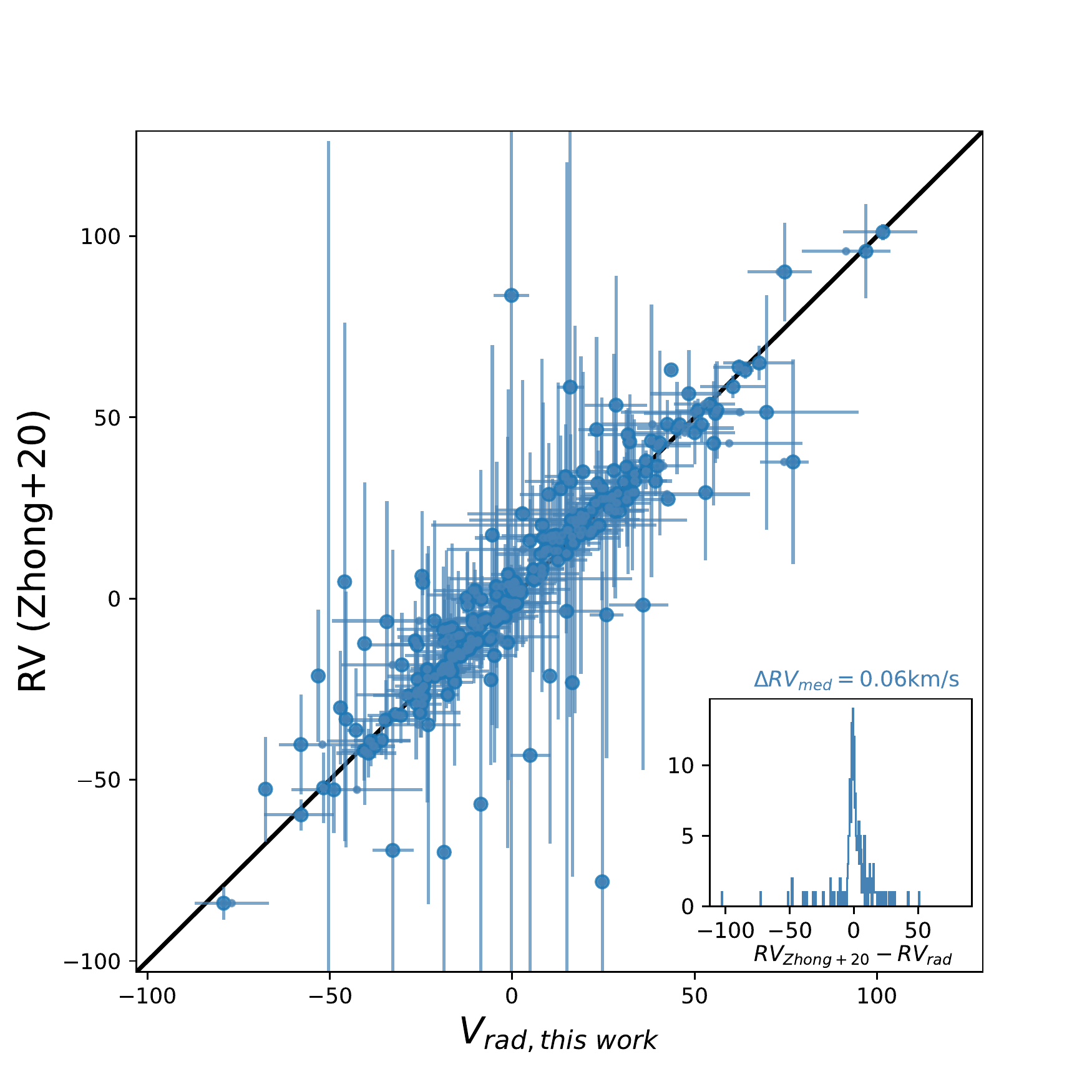}
    \end{subfigure}
    \caption{Comparison of our cluster $V_{\rm rad}$ to results in the
    literature. The left, middle, and right panels show comparison to 308,
    185, and 226 OCs of \citet{Tarricq2021},  \citet{Dias2021}, and
    \citet{Zhong2020}, respectively.  The histograms of the cluster radial
    velocity differences are shown as insets in each panel.  }
    \label{fig:knownrv}
\end{figure*}

Among all our 386 OCs, we found 308 in common with \citet{Tarricq2021}, 185 in
common with \citet{Dias2021}, and 226 in common with \citet{Zhong2020}.
Figure~\ref{fig:knownrv} shows the comparison of the cluster $V_{\rm rad}$
between our results and these three catalogues. The distributions of the
$V_{\rm rad}$ difference are also displayed as a histogram in each panel.
Compared to the previous LAMOST OC results from \citet{Zhong2020}, the median
value of the $V_{\rm rad}$ difference is only 0.06 km~s$^{-1}$(see the right
panel of Fig.  \ref{fig:knownrv} ), which indicates that the two catalogues are
consistent in $V_{\rm rad}$.  The median $V_{\rm rad}$ difference between our
results and the ones from \citet{Tarricq2021} and \citet{Dias2021} have a value
around 6.2 km~s$^{-1}$ (see the left and middle panel of Fig.
\ref{fig:knownrv}).  These differences should, at least partially, be due to
the aforementioned systematic offset of LAMOST RV.

In total, 342 of the OCs in our catalogue already have a $V_{\rm rad}$ measurement in one of these three recent large catalogues. Their $V_{\rm
rad}$, $\sigma\,V_{\rm rad}$, number of stars used in the calculation, together
with the literature values, are listed in Table \ref{tab:knownrv}. 

\begin{table*}
\centering
\setlength{\tabcolsep}{1mm}
\caption{Radial velocity of 342 OCs with known $V_{\rm rad}$ in the literature.}
\begin{tabular}{l|rrrrr|rr|rr|rr}
\hline
\multicolumn{1}{c}{cluster}  &
\multicolumn{1}{c}{N$_{all}$}  &
\multicolumn{1}{c}{N$_{flag1}$}  &
\multicolumn{1}{c}{V$_{rad, med}$}  &
\multicolumn{1}{c}{V$_{rad, mean}$}  &
\multicolumn{1}{c}{$\sigma$V$_{rad}$}  &
\multicolumn{1}{c}{RV$_{Tarricq21}$} &     
\multicolumn{1}{c}{N$_{RV}$ }  &     
\multicolumn{1}{c}{RV$_{Dias21}$} &     
\multicolumn{1}{c}{N$_{RV}$ }  &     
\multicolumn{1}{c}{RV$_{Zhong20}$} &
\multicolumn{1}{c}{N$_{RV}$ }      \\
\hline
 ASCC\_10   & 23   & 22  &   -24.20 $\pm$ 1.57  & -24.58 $\pm$ 1.23  & 7.42   & -15.61  $\pm$ 4.35  & 3    &    -10.87 $\pm$ 5.24  & 3  & -27.21 $\pm$ 5.91   & 33  \\
 ASCC\_11   & 39   & 39  &   -21.37 $\pm$ 1.94  & -21.02 $\pm$ 1.51  & 13.18  & -14.44  $\pm$ 0.32  & 3    &    -14.24 $\pm$ 0.23  & 3  & -21.49 $\pm$ 12.74  & 52  \\
 ASCC\_105  & 5    & 5   &   -18.98 $\pm$ 3.05  & -18.47 $\pm$ 3.16  & 5.95   & -6.93   $\pm$ 1.36  & 19   &   -14.30  $\pm$ 4.09  & 18 & -20.01 $\pm$ 2.47   & 6   \\
 ASCC\_108  & 11   & 11  &   -14.93 $\pm$ 3.69  & -14.75 $\pm$ 3.41  & 10.84  & -29.67  $\pm$ 19.60 & 2    &                       &    & -14.73 $\pm$ 6.22   & 12  \\
 ...        &...   &...  &...                   &...                 &...     &...                  &...   &...                    &... &...                  &   \\
\hline
\end{tabular}
\label{tab:knownrv}
\end{table*}

\subsubsection{Clusters with newly-obtained $V_{\rm rad}$}
\label{subsec:newrv}

The large number of LAMOST spectra enables us to obtain cluster $V_{\rm rad}$
values that were not reported in the literature before. We report here 44
clusters with newly-obtained  $V_{\rm rad}$.  Table \ref{tab:newrv} lists their
information. These values, together with those listed in
Table~\ref{tab:knownrv}, will be employed later in the paper, for a kinematical
analysis of the OC population.

\begin{table}
\centering
\setlength{\tabcolsep}{0.6mm}
\caption{Newly-obtained cluster radial velocity of 44 OCs.}
\begin{tabular}{lrrrrr}
\hline
\multicolumn{1}{c}{cluster}  &
\multicolumn{1}{c}{N$_{all}$}  &
\multicolumn{1}{c}{N$_{flag1}$}  &
\multicolumn{1}{c}{V$_{rad,med}$}  &
\multicolumn{1}{c}{V$_{rad,mean}$}  &
\multicolumn{1}{c}{$\sigma$V$_{rad}$}  \\
\hline
 COIN-Gaia\_24 & 44 & 44 & -9.20  $\pm$ 1.35  & -9.20  $\pm$ 1.16  & 8.22  \\
 LP\_2139      & 36 & 34 & -4.34  $\pm$ 2.06  & -4.03  $\pm$ 1.67  & 11.62  \\
 UBC\_88       & 29 & 28 & -15.78 $\pm$ 2.12  & -16.76 $\pm$ 1.71  & 12.11  \\
 COIN-Gaia\_23 & 18 & 17 & 1.72   $\pm$ 3.44  & 1.35   $\pm$ 2.71  & 12.55  \\
 COIN-Gaia\_19 & 17 & 17 & 8.53   $\pm$ 2.70  & 8.81   $\pm$ 2.75  & 12.64  \\
 UBC\_616      & 13 & 13 & 52.19  $\pm$ 3.56  & 50.51  $\pm$ 2.53  & 14.18  \\
 COIN-Gaia\_17 & 13 & 12 & 6.80   $\pm$ 2.97  & 6.84   $\pm$ 2.38  & 8.94  \\
 UPK\_282      & 11 & 11 & -16.46 $\pm$ 2.39  & -16.80 $\pm$ 1.92  & 7.94  \\
 UBC\_51       & 9  & 9  & -14.65 $\pm$ 4.73  & -19.19 $\pm$ 3.48  & 17.67  \\
 UBC\_56       & 8  & 8  & -12.43 $\pm$ 4.44  & -11.97 $\pm$ 3.87  & 10.75  \\
 COIN-Gaia\_14 & 8  & 8  & 4.56   $\pm$ 3.46  & 5.17   $\pm$ 2.66  & 10.80  \\
 COIN-Gaia\_39 & 7  & 7  & -3.84  $\pm$ 4.86  & -0.49  $\pm$ 3.70  & 15.47  \\
 UPK\_119      & 5  & 5  & -24.34 $\pm$ 3.16  & -24.37 $\pm$ 2.55  & 5.30  \\
 UBC\_63       & 5  & 5  & -22.94 $\pm$ 4.49  & -22.15 $\pm$ 3.50  & 8.99  \\
 UBC\_80       & 4  & 4  & 38.54  $\pm$ 8.40  & 38.92  $\pm$ 7.91  & 13.37  \\
 UBC\_615      & 4  & 4  & 13.47  $\pm$ 8.18  & 13.32  $\pm$ 5.87  & 25.94  \\
 COIN-Gaia\_20 & 4  & 4  & -4.35  $\pm$ 5.00  & -4.97  $\pm$ 5.06  & 8.66  \\
 UBC\_201      & 4  & 4  & 19.08  $\pm$ 6.98  & 19.52  $\pm$ 6.14  & 12.94  \\
 UBC\_188      & 4  & 4  & -35.56 $\pm$ 6.49  & -36.03 $\pm$ 7.86  & 12.79  \\
 UPK\_312      & 4  & 4  & -12.03 $\pm$ 4.01  & -11.75 $\pm$ 3.79  & 6.72  \\
 UBC\_77       & 3  & 3  & 17.30  $\pm$ 5.95  & 17.67  $\pm$ 5.26  & 7.00  \\
 UBC\_68       & 3  & 3  & 13.19  $\pm$ 5.87  & 13.06  $\pm$ 4.78  & 7.31  \\
 NGC\_381      & 3  & 3  & -22.88 $\pm$ 6.32  & -22.93 $\pm$ 5.57  & 7.06  \\
 COIN-Gaia\_40 & 3  & 3  & 7.47   $\pm$ 7.92  & 8.42   $\pm$ 7.08  & 11.55  \\
 UBC\_437      & 3  & 3  & 19.37  $\pm$ 10.24 & 18.60  $\pm$ 7.24  & 15.07  \\
 UBC\_417      & 3  & 3  & -47.45 $\pm$ 7.35  & -47.74 $\pm$ 4.63  & 9.88  \\
 NGC\_2169     & 3  & 3  & 14.32  $\pm$ 4.22  & 14.47  $\pm$ 3.67  & 4.93  \\
 SAI\_14       & 2  & 2  & -69.73 $\pm$ 3.00  & -69.73 $\pm$ 3.00  & 2.51  \\
 UBC\_198      & 2  & 2  & -7.13  $\pm$ 7.70  & -7.13  $\pm$ 7.70  & 7.18  \\
 UBC\_129      & 2  & 2  & -17.90 $\pm$ 7.33  & -17.90 $\pm$ 7.33  & 6.08  \\
 UBC\_216      & 1  & 1  & -1.11  $\pm$ 10.05 &                    & \\
 UBC\_395      & 1  & 1  & -13.62 $\pm$ 17.02 &                    & \\
 UBC\_150      & 1  & 1  & -17.43 $\pm$ 6.96  &                    & \\
 Teutsch\_8    & 1  & 1  & -17.62 $\pm$ 6.45  &                    & \\
 UPK\_131      & 1  & 1  & -20.25 $\pm$ 5.84  &                    & \\
 UBC\_182      & 1  & 1  & -28.86 $\pm$ 5.37  &                    & \\
 UBC\_596      & 1  & 1  & -29.18 $\pm$ 15.03 &                    & \\
 UBC\_430      & 1  & 1  & -33.62 $\pm$ 12.74 &                    & \\
 UBC\_421      & 1  & 1  & -40.23 $\pm$ 13.84 &                    & \\
 UBC\_214      & 1  & 1  & 42.81  $\pm$ 11.46 &                    & \\
 UBC\_442      & 1  & 1  & 30.82  $\pm$ 5.93  &                    & \\
 UBC\_206      & 1  & 1  & 27.55  $\pm$ 10.13 &                    & \\
 COIN-Gaia\_21 & 1  & 1  & 1.41   $\pm$ 10.77 &                    & \\
 UBC\_435      & 1  & 1  & -0.68  $\pm$ 13.71 &                    & \\
\hline
\end{tabular}
\label{tab:newrv}
\end{table}

\subsection{Metallicity}
\label{sec:feh}

As groups of stars form together with the same composition, OCs are an ideal
probe to trace the metallicity evolution of the Galactic disc.  
With OCs, in fact, not only we can improve the S/N by averaging over several member
stars, but also we could deal with objects whose age can be determined with a
lower uncertainty than that of field stars (with the only possible exception of
stars for which a full asteroseismologic analysis is possible, see
e.g. \citet{Rodrigues2017, Miglio2021} ).  

In this section we report the metallicity of LAMOST OCs, which consists of 218
clusters with known metallicity collected from the literature, and 137 clusters
with newly-obtained spectroscopic metallicity. Here we use the iron abundance
\feh as an index of the stellar metallicity.


\subsubsection{Clusters with \feh in the literature}
\label{subsec:knownfeh}

\begin{figure*}
    \centering
    \begin{subfigure}[]{0.48\textwidth}
        \includegraphics[width=\textwidth]{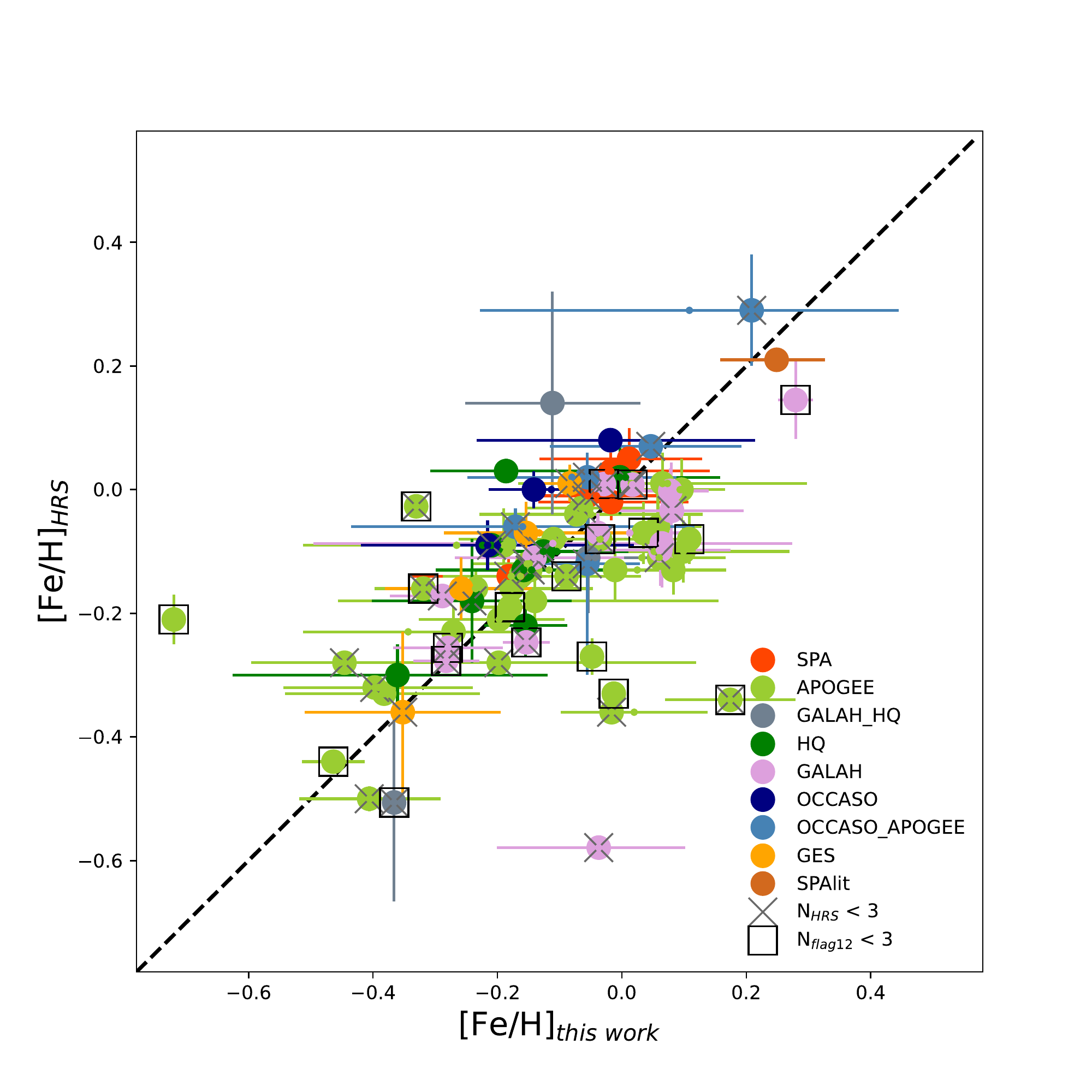}
    \end{subfigure}
    \begin{subfigure}[]{0.48\textwidth}
        \includegraphics[width=\textwidth]{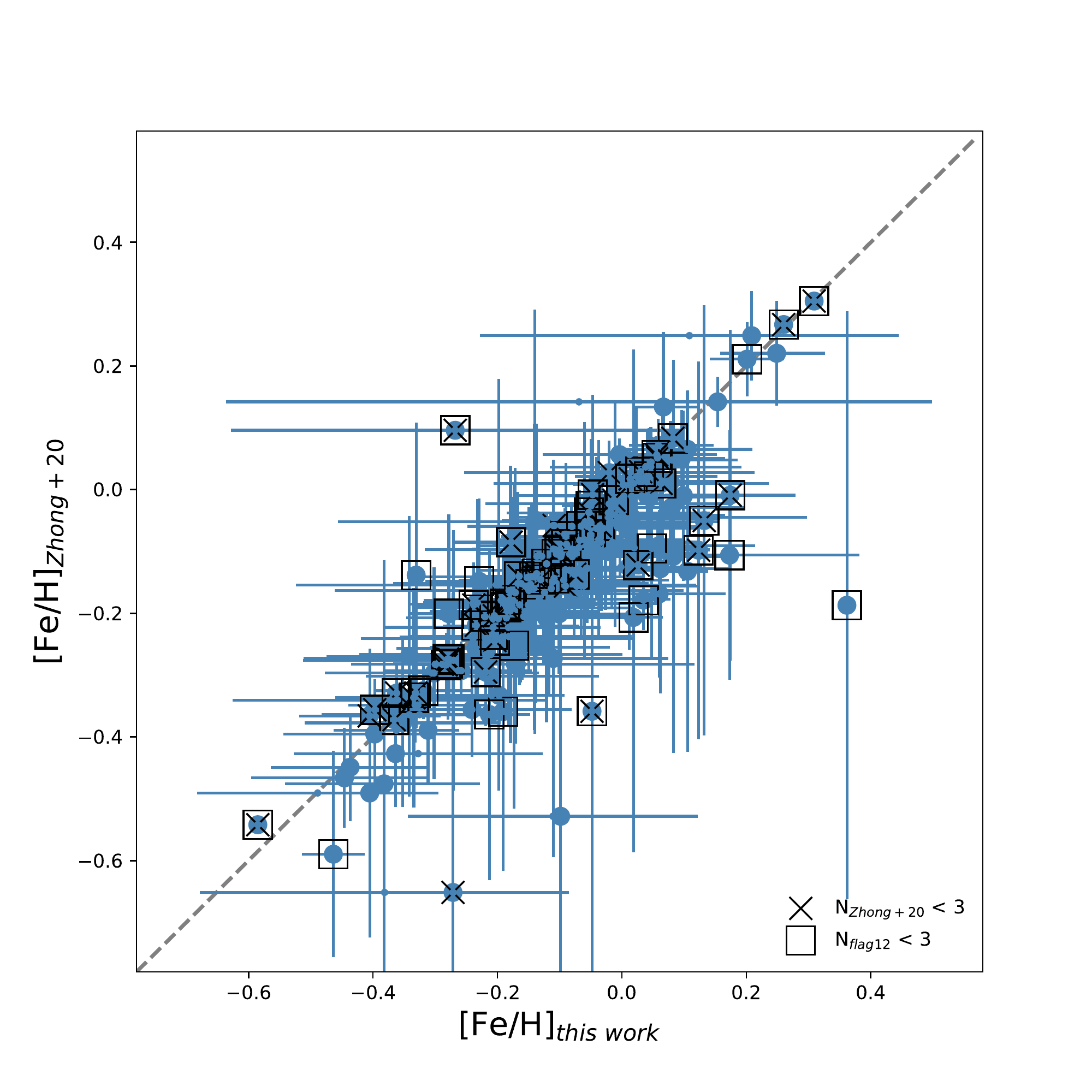}
    \end{subfigure}
    \caption{Comparison of 82 clusters' \feh to literature results with high resolution spectra (HRS) in the left panel, and of 206 clusters to results from \citet{Zhong2020}.
    The symbol X marks \feh values with less than 3 member stars, namely 38 clusters on the left (from HRS) and 75 clusters on the right \citep[from ][]{Zhong2020}.   
    The open square marks clusters with less than three {\sc flag=12} member stars from this work, which are 23 clusters on the left panel and 92 clusters on the right panel.   
    }
    \label{fig:feh}
\end{figure*}

We collect cluster \feh measurements from high resolution (HRS) spectra
including results of large surveys such as $Gaia$-ESO \citep[][and references
therein]{Spina2017, Bragaglia2021,Randich2022}, APOGEE \citep{Donor2020,
Carrera2019, Spina2021}, GALAH \citep{Carrera2019, Spina2021}, large projects
like SPA \citep{Origlia2019, Casali2020b, Zhang2021, AlonsoSantiago2021},
OCCASO \citep{Casamiquela2016, Casamiquela2017, Casamiquela2018}, and high
quality (HQ) collections from \citet{Netopil2022}. In general, the HRS spectra
results we use are similar to the one used in \citet{Zhang2021}, but with more
OCs from \citet{Spina2021} and HQ results of \citet{Netopil2022}. In total, we
have 82 clusters in common with HRS \feh results. The left panel of Fig.
\ref{fig:feh} shows the comparison between our results based on LAMOST and the
HRS values. The difference between the cluster \feh values of these two data
sets, is 0.01$\pm$0.15  dex.

Clusters with less than three member stars for the \feh determination, both in
our case (marked with open square of Fig. \ref{fig:feh}) and in the HRS works
(marked with X), show a relatively large difference between the two sets of
result. If we compare clusters with at least three \feh member stars in both
datasets, their difference is 0.00$\pm$0.11 dex, which shows a very good
consistency. Therefore, in the following discussions on the Galactic
metallicity evolution we only consider clusters with at least three members
and with {\sc flag=12}.

We also compare our results to the previous LAMOST OC work by \citet{Zhong2020}
which is based on an earlier data release and different member star selection.
There are 206 clusters in the \citet{Zhong2020} catalogue with \feh
determination, the right panel of Fig. \ref{fig:feh} shows the \feh
comparison. For all the matched clusters, the \feh value mean difference is
0.02$\pm$0.23 dex. When only clusters with at least three \feh member stars are
considered, their difference is 0.04$\pm$0.06 dex.

Combining both literature results from HRS studies and low resolution spectra
study of \citet{Zhong2020}, 218 of our clusters have a reported metallicity.  Table \ref{tab:knownfeh} lists their information.

\begin{table*}
\centering
\setlength{\tabcolsep}{.7mm}
\caption{218 clusters with [Fe/H] values in literature.}
\begin{tabular}{l|rrrr|rrr|rr}
\hline
\multicolumn{1}{c}{cluster}  &
\multicolumn{1}{c}{N$_{flag=12}$}  &
\multicolumn{1}{c}{[Fe/H]$_{med}$}  &
\multicolumn{1}{c}{[Fe/H]$_{mean}$}  &
\multicolumn{1}{c}{$\sigma$[Fe/H]}  &
\multicolumn{1}{c}{[Fe/H]$_{HRS}$} &     
\multicolumn{1}{c}{N$_{HRS}$ }  &     
\multicolumn{1}{c}{source} &     
\multicolumn{1}{c}{[Fe/H]$_{Zhong20}$} &     
\multicolumn{1}{c}{N$_{Zhong20}$ }      \\
\hline
ASCC\_10 &  12 & -0.03 $\pm$ 0.01 & -0.02 $\pm$ 0.01 & 0.06 &                  &   &     &  -0.10 $\pm$ 0.10  & 22  \\
ASCC\_11 &  28 & -0.18 $\pm$ 0.03 & -0.16 $\pm$ 0.03 & 0.18 & -0.14 $\pm$ 0.05 & 1 & SPA &  -0.24 $\pm$ 0.09  & 26 \\
 ...&  ...&  ...&  ...&  ...&  ...&  ...&  ...&  ...& ...\\
\hline
\end{tabular}
\label{tab:knownfeh}
\end{table*}

\subsubsection{Clusters with newly-obtained \feh}
\label{subsec:newfeh}

Here we report \feh measurements of 137 OCs which do not have spectroscopic \feh in the literature.
Among them, 63 clusters have at least three {\sc flag=12} member stars.
Table \ref{tab:newfeh3} and table \ref{tab:newfeh21} list our results for these clusters. 
Clusters with less than three {\sc flag=12} member stars are available in our final catalog as well, but their metallicities should be used with caution.
This constitutes an important addition to the number of OCs with metallicity determined on the basis of spectroscopy.

\begin{table*}
\centering
\setlength{\tabcolsep}{1mm}
\caption{Newly-obtained iron abundances for 63 OCs with at least three {\sc flag=12} member stars.}
\begin{tabular}{lrrrr|| lrrrr}
\hline
\multicolumn{1}{c}{cluster}  &
\multicolumn{1}{c}{N$_{flag12}$}  &
\multicolumn{1}{c}{[Fe/H]$_{med}$}  &
\multicolumn{1}{c}{[Fe/H]$_{mean}$}  &
\multicolumn{1}{c}{$\sigma$[Fe/H]}  &
\multicolumn{1}{c}{cluster}  &
\multicolumn{1}{c}{N$_{flag12}$}  &
\multicolumn{1}{c}{[Fe/H]$_{med}$}  &
\multicolumn{1}{c}{[Fe/H]$_{mean}$}  &
\multicolumn{1}{c}{$\sigma$[Fe/H]}  \\
\hline
 COIN-Gaia\_9   & 4  & 0.01   $\pm$  0.03 & -0.01 $\pm$ 0.02 & 0.09    &    UBC\_6        & 4  & -0.01  $\pm$  0.02 & 0.00  $\pm$ 0.02 & 0.06  \\
 COIN-Gaia\_10  & 3  & -0.06  $\pm$  0.11 & -0.08 $\pm$ 0.09 & 0.16    &    UBC\_8        & 38 & -0.12  $\pm$  0.02 & -0.12 $\pm$ 0.01 & 0.12  \\
 COIN-Gaia\_11  & 23 & 0.16   $\pm$  0.03 & 0.16  $\pm$ 0.02 & 0.18    &    UBC\_51       & 3  & 0.04   $\pm$  0.02 & 0.05  $\pm$ 0.03 & 0.03  \\
 COIN-Gaia\_12  & 7  & -0.00  $\pm$  0.03 & -0.03 $\pm$ 0.05 & 0.14    &    UBC\_53       & 4  & -0.04  $\pm$  0.04 & -0.04 $\pm$ 0.03 & 0.07  \\
 COIN-Gaia\_13  & 30 & -0.07  $\pm$  0.01 & -0.05 $\pm$ 0.02 & 0.13    &    UBC\_54       & 37 & -0.08  $\pm$  0.02 & -0.08 $\pm$ 0.02 & 0.16  \\
 COIN-Gaia\_14  & 6  & 0.01   $\pm$  0.08 & 0.01  $\pm$ 0.07 & 0.17    &    UBC\_55       & 11 & -0.12  $\pm$  0.03 & -0.11 $\pm$ 0.03 & 0.12  \\
 COIN-Gaia\_17  & 11 & -0.07  $\pm$  0.05 & -0.02 $\pm$ 0.03 & 0.18    &    UBC\_56       & 3  & -0.05  $\pm$  0.11 & -0.02 $\pm$ 0.11 & 0.17  \\
 COIN-Gaia\_18  & 17 & -0.07  $\pm$  0.03 & -0.08 $\pm$ 0.03 & 0.16    &    UBC\_59       & 3  & -0.11  $\pm$  0.04 & -0.13 $\pm$ 0.08 & 0.10  \\
 COIN-Gaia\_19  & 12 & -0.06  $\pm$  0.03 & -0.09 $\pm$ 0.04 & 0.15    &    UBC\_63       & 3  & -0.12  $\pm$  0.08 & -0.13 $\pm$ 0.08 & 0.11  \\
 COIN-Gaia\_20  & 3  & -0.06  $\pm$  0.04 & -0.06 $\pm$ 0.02 & 0.05    &    UBC\_74       & 33 & -0.10  $\pm$  0.01 & -0.09 $\pm$ 0.02 & 0.10  \\
 COIN-Gaia\_23  & 12 & -0.16  $\pm$  0.05 & -0.14 $\pm$ 0.06 & 0.21    &    UBC\_77       & 3  & -0.04  $\pm$  0.04 & -0.05 $\pm$ 0.05 & 0.06  \\
 COIN-Gaia\_24  & 42 & 0.06   $\pm$  0.01 & 0.01  $\pm$ 0.01 & 0.13    &    UBC\_88       & 20 & -0.02  $\pm$  0.02 & -0.04 $\pm$ 0.02 & 0.13  \\
 COIN-Gaia\_26  & 9  & -0.13  $\pm$  0.04 & -0.11 $\pm$ 0.04 & 0.21    &    UBC\_434      & 3  & -0.46  $\pm$  0.07 & -0.47 $\pm$ 0.06 & 0.08  \\
 COIN-Gaia\_27  & 3  & -0.03  $\pm$  0.08 & -0.03 $\pm$ 0.08 & 0.10    &    UBC\_586      & 3  & 0.08   $\pm$  0.07 & 0.08  $\pm$ 0.05 & 0.08  \\
 COIN-Gaia\_28  & 3  & -0.14  $\pm$  0.05 & -0.13 $\pm$ 0.06 & 0.07    &    UBC\_609      & 3  & -0.11  $\pm$  0.09 & -0.12 $\pm$ 0.07 & 0.13  \\
 COIN-Gaia\_38  & 5  & 0.02   $\pm$  0.10 & 0.03  $\pm$ 0.07 & 0.19    &    UBC\_614      & 3  & -0.01  $\pm$  0.11 & -0.00 $\pm$ 0.07 & 0.14  \\
 COIN-Gaia\_39  & 3  & -0.08  $\pm$  0.10 & -0.04 $\pm$ 0.10 & 0.17    &    UBC\_615      & 4  & -0.15  $\pm$  0.14 & -0.15 $\pm$ 0.11 & 0.23  \\
 COIN-Gaia\_41  & 5  & -0.14  $\pm$  0.07 & -0.14 $\pm$ 0.07 & 0.17    &    UBC\_616      & 4  & -0.25  $\pm$  0.04 & -0.28 $\pm$ 0.05 & 0.14  \\
 LP\_2139      & 30 & -0.19  $\pm$  0.03 & -0.17 $\pm$ 0.03 & 0.19     &    UBC\_619      & 5  & -0.18  $\pm$  0.06 & -0.17 $\pm$ 0.08 & 0.15  \\
 NGC\_381      & 3  & -0.13  $\pm$  0.05 & -0.10 $\pm$ 0.03 & 0.10     &    UBC\_622      & 3  & -0.15  $\pm$  0.03 & -0.15 $\pm$ 0.02 & 0.05  \\
 NGC\_1502     & 5  & 0.04   $\pm$  0.06 & 0.03  $\pm$ 0.04 & 0.10     &    UPK\_119      & 5  & 0.03   $\pm$  0.02 & 0.03  $\pm$ 0.02 & 0.08  \\
 NGC\_2126     & 21 & -0.23  $\pm$  0.02 & -0.21 $\pm$ 0.03 & 0.17     &    UPK\_166      & 7  & 0.02   $\pm$  0.03 & -0.07 $\pm$ 0.08 & 0.26  \\
 NGC\_6997     & 3  & 0.07   $\pm$  0.04 & 0.06  $\pm$ 0.05 & 0.06     &    UPK\_168      & 4  & 0.10   $\pm$  0.02 & 0.07  $\pm$ 0.05 & 0.10  \\
 UBC\_2        & 7  & -0.05  $\pm$  0.04 & -0.04 $\pm$ 0.05 & 0.12     &    UPK\_185      & 10 & 0.01   $\pm$  0.01 & 0.01  $\pm$ 0.01 & 0.04  \\
 UBC\_4        & 14 & -0.11  $\pm$  0.02 & -0.08 $\pm$ 0.01 & 0.15     &    UPK\_282      & 7  & -0.05  $\pm$  0.03 & -0.06 $\pm$ 0.03 & 0.08  \\
 UBC\_13       & 19 & -0.07  $\pm$  0.03 & -0.05 $\pm$ 0.03 & 0.13     &    UPK\_294      & 3  & -0.16  $\pm$  0.07 & -0.15 $\pm$ 0.06 & 0.09  \\
 UBC\_19       & 7  & 0.09   $\pm$  0.01 & 0.08  $\pm$ 0.05 & 0.10     &    UPK\_296      & 16 & -0.05  $\pm$  0.01 & -0.03 $\pm$ 0.02 & 0.14  \\
 UBC\_31       & 23 & -0.01  $\pm$  0.01 & 0.00  $\pm$ 0.01 & 0.08     &    UPK\_305      & 12 & 0.05   $\pm$  0.02 & 0.05  $\pm$ 0.01 & 0.07  \\
 UBC\_188      & 3  & 0.02   $\pm$  0.04 & -0.17 $\pm$ 0.05 & 0.31     &    UPK\_350      & 13 & -0.06  $\pm$  0.03 & -0.04 $\pm$ 0.03 & 0.12  \\
 UBC\_200      & 17 & -0.21  $\pm$  0.03 & -0.16 $\pm$ 0.03 & 0.21     &    UPK\_381      & 8  & -0.04  $\pm$  0.08 & -0.04 $\pm$ 0.06 & 0.18  \\
 UBC\_203      & 6  & -0.39  $\pm$  0.05 & -0.37 $\pm$ 0.03 & 0.14     &    UPK\_429      & 9  & 0.01   $\pm$  0.05 & 0.02  $\pm$ 0.07 & 0.19  \\
 UBC\_433      & 3  & -0.52  $\pm$  0.13 & -0.52 $\pm$ 0.10 & 0.17     &     &&&&  \\
\hline
\end{tabular}
\label{tab:newfeh3}
\end{table*}

\begin{table*}
\centering
\setlength{\tabcolsep}{1.mm}
\caption{Newly-obtained iron abundances for 74 OCs but with less than three {\sc flag=12} member stars.}
\begin{tabular}{lrrrr || lrrrr}
\hline
\multicolumn{1}{c}{cluster}  &
\multicolumn{1}{c}{N$_{flag12}$}  &
\multicolumn{1}{c}{[Fe/H]$_{med}$}  &
\multicolumn{1}{c}{[Fe/H]$_{mean}$}  &
\multicolumn{1}{c}{$\sigma$[Fe/H]}  &
\multicolumn{1}{c}{cluster}  &
\multicolumn{1}{c}{N$_{flag12}$}  &
\multicolumn{1}{c}{[Fe/H]$_{med}$}  &
\multicolumn{1}{c}{[Fe/H]$_{mean}$}  &
\multicolumn{1}{c}{$\sigma$[Fe/H]}  \\
\hline
 Berkeley\_34   & 2 & -0.21 $\pm$  0.06 & -0.21 $\pm$  0.06 & 0.05     &      UBC\_216       & 1 & -0.07 $\pm$  0.12 &                   &       \\
 COIN-Gaia\_2   & 1 & -0.24 $\pm$  0.21 &                   &          &      UBC\_374       & 2 & 0.19  $\pm$  0.03 & 0.19  $\pm$  0.03 & 0.03  \\
 COIN-Gaia\_8   & 2 & -0.05 $\pm$  0.04 & -0.05 $\pm$  0.04 & 0.08     &      UBC\_395       & 1 & -0.06 $\pm$  0.09 &                   &       \\
 COIN-Gaia\_15  & 1 & -0.01 $\pm$  0.09 &                   &          &      UBC\_417       & 2 & -0.04 $\pm$  0.02 & -0.04 $\pm$  0.02 & 0.05  \\
 COIN-Gaia\_16  & 1 & -0.12 $\pm$  0.08 &                   &          &      UBC\_419       & 1 & -0.09 $\pm$  0.06 &                   &       \\
 COIN-Gaia\_21  & 1 & -0.11 $\pm$  0.19 &                   &          &      UBC\_421       & 1 & -0.37 $\pm$  0.19 &                   &       \\
 COIN-Gaia\_22  & 1 & -0.26 $\pm$  0.26 &                   &          &      UBC\_427       & 1 & 0.20  $\pm$  0.06 &                   &       \\
 Collinder\_115 & 1 & -0.17 $\pm$  0.19 &                   &          &      UBC\_428       & 2 & -0.35 $\pm$  0.16 & -0.35 $\pm$  0.16 & 0.13  \\
 Collinder\_421 & 2 & -0.02 $\pm$  0.04 & -0.02 $\pm$  0.04 & 0.04     &      UBC\_430       & 1 & -0.45 $\pm$  0.13 &                  &      \\
 Czernik\_38    & 1 & 0.54  $\pm$  0.10 &                   &          &      UBC\_431       & 1 & -0.47 $\pm$  0.15 &                  &      \\
 Dolidze\_5     & 1 & -1.93 $\pm$  0.16 &                   &          &      UBC\_436       & 2 & -0.19 $\pm$  0.04 & -0.19 $\pm$ 0.04 & 0.03  \\
 FSR\_0932      & 1 & -2.17 $\pm$  0.06 &                   &          &      UBC\_437       & 2 & -0.20 $\pm$  0.14 & -0.20 $\pm$ 0.14 & 0.12  \\
 FSR\_0975      & 1 & -0.18 $\pm$  0.22 &                   &          &      UBC\_438       & 1 & -0.39 $\pm$  0.06 &                  &      \\
 LP\_658        & 1 & -0.32 $\pm$  0.08 & -0.32 $\pm$  0.08 & 0.00     &      UBC\_440       & 2 & -0.26 $\pm$  0.05 & -0.26 $\pm$ 0.05 & 0.10  \\
 LP\_930        & 1 & 0.16  $\pm$  0.04 &                   &          &      UBC\_442       & 1 & -0.08 $\pm$  0.06 &                  &      \\
 LP\_2198       & 2 & 0.25  $\pm$  0.15 & 0.25  $\pm$  0.15 & 0.13     &      UBC\_445       & 1 & -0.04 $\pm$  0.33 &                  &      \\
 NGC\_457       & 1 & -0.14 $\pm$  0.18 &                   &          &      UBC\_587       & 1 & -0.22 $\pm$  0.08 &                  &      \\
 NGC\_2169      & 1 & -0.05 $\pm$  0.04 &                   &          &      UBC\_596       & 1 & -0.19 $\pm$  0.07 &                  &      \\
 NGC\_6871      & 1 & 0.26  $\pm$  0.03 &                   &          &      UBC\_607       & 2 & -0.38 $\pm$  0.09 & -0.38 $\pm$ 0.09 & 0.12  \\
 SAI\_14        & 2 & -0.41 $\pm$  0.08 & -0.41 $\pm$  0.08 & 0.07     &      UBC\_610       & 2 & -0.26 $\pm$  0.06 & -0.26 $\pm$ 0.06 & 0.23  \\
 Sigma\_Ori     & 2 & -0.46 $\pm$  0.06 & -0.46 $\pm$  0.06 & 0.22     &      UBC\_629       & 2 & -0.38 $\pm$  0.06 & -0.38 $\pm$ 0.06 & 0.04  \\
 Teutsch\_8     & 1 & 0.11  $\pm$  0.18 &                   &          &      UPK\_45        & 2 & 0.08  $\pm$  0.06 & 0.08  $\pm$ 0.06 & 0.05  \\
 Teutsch\_35    & 2 & 0.12  $\pm$  0.03 & 0.12  $\pm$  0.03 & 0.03     &      UPK\_65        & 1 & 0.02  $\pm$  0.13 &                  &      \\
 UBC\_49        & 2 & -0.18 $\pm$  0.06 & -0.18 $\pm$ 0.06  & 0.08     &      UPK\_79        & 2 & -0.04 $\pm$  0.04 & -0.04 $\pm$ 0.04 & 0.04  \\
 UBC\_52        & 1 & 0.04  $\pm$  0.08 &                   &          &      UPK\_82        & 1 & 0.04  $\pm$  0.01 &                  &      \\
 UBC\_57        & 2 & -0.25 $\pm$  0.09 & -0.25 $\pm$ 0.09  & 0.07     &      UPK\_93        & 2 & -0.01 $\pm$  0.01 & -0.01 $\pm$ 0.01 & 0.01  \\
 UBC\_61        & 1 & 0.10  $\pm$  0.16 &                  &           &      UPK\_108       & 2 & 0.00  $\pm$  0.02 & 0.00  $\pm$ 0.02 & 0.03  \\
 UBC\_68        & 2 & -0.12 $\pm$  0.10 & -0.12 $\pm$ 0.10 & 0.09      &      UPK\_131       & 1 & 0.02  $\pm$  0.03 &                  &      \\
 UBC\_73        & 2 & -0.19 $\pm$  0.18 & -0.19 $\pm$ 0.18 & 0.14      &      UPK\_136       & 2 & -0.05 $\pm$  0.05 & -0.05 $\pm$ 0.05 & 0.14  \\
 UBC\_82        & 1 & -0.13 $\pm$  0.03 &                  &           &      UPK\_303       & 2 & 0.12  $\pm$  0.02 & 0.12  $\pm$ 0.02 & 0.02  \\
 UBC\_90        & 1 & -0.28 $\pm$  0.15 & -0.28 $\pm$ 0.15 & 0.00      &      UPK\_333       & 2 & -0.05 $\pm$  0.05 & -0.05 $\pm$ 0.05 & 0.05  \\
 UBC\_141       & 2 & -0.09 $\pm$  0.04 & -0.09 $\pm$  0.04 & 0.03     &      UPK\_369       & 2 & 0.18  $\pm$  0.03 & 0.18  $\pm$ 0.03 & 0.20  \\
 UBC\_150       & 1 & -0.05 $\pm$  0.01 &                   &          &      UPK\_379       & 1 & -0.05 $\pm$  0.03 &                  &       \\
 UBC\_169       & 1 & -0.08 $\pm$  0.02 &                   &          &      UPK\_385       & 2 & 0.13  $\pm$  0.01 & 0.13  $\pm$ 0.01 & 0.04  \\
 UBC\_176       & 1 & 0.12  $\pm$  0.21 &                   &          &      UPK\_418       & 2 & -0.25 $\pm$  0.07 & -0.25 $\pm$ 0.07 & 0.11  \\
 UBC\_182       & 1 & -0.14 $\pm$  0.29 &                   &          &      UPK\_422       & 1 & 0.06  $\pm$  0.06 &                  &      \\
 UBC\_197       & 1 & -0.08 $\pm$  0.03 &                   &          &      vdBergh\_85    & 1 & -0.24 $\pm$  0.11 &                  &      \\
\hline
\end{tabular}
\label{tab:newfeh21}
\end{table*}

\section{Discussions}
\label{sec:dis}

\subsection{The Galactic metallicity distribution and dynamical properties of the LAMOST OCs}
\label{sec:grad}

In order to study the metallicity distribution of the Milky Way disc, one of the most
common subjects investigated is the so-called radial metallicity gradient.
The metallicity gradient is often displayed with \feh as a function of the 
Galactocentric radius \rgc, indicating different levels of metal enrichment along the Galactic radius.
Stellar open clusters, being groups of stars in a simple stellar population,
are ideal probes to trace the metal enrichment history of the Milky Way.

With the homogeneous V${\sc_ {rad}}$ and \feh determination in our newly compiled LAMOST OC catalogue,
we can investigate the evolution of the Galactic metallicity --  even its gradient evolution -- in the past 500 Myr. 
This 500-Myr range traces back to the time when the Sagittarius dwarf galaxy (Sgr
dSph) had its last passage through the MW outer disc \citep[see for instance
the discussions in][]{Xu2020}. Sgr dSph is also the last relatively
considerable minor merger in the evolution history of our galaxy known to date,
which means the gravitational potential of the Galaxy can be considered as a
constant in the past 500 Myr. A stable gravitational potential is further the
guarantee of orbit calculations for these clusters.

To trace back the orbits of LAMOST OCs and their evolution, we use the publicly
licensed code \texttt{GALPOT} \footnote{\url{https://github.com/PaulMcMillan-Astro/GalPot }}
\citep{Dehnen1998} and the Milky Way gravitational potential from
\citet{McMillan2017}.  The gravitational potential takes into account the
Galactic thick and thin stellar discs, a bulge component, a dark-matter halo,
and a cold gas disc. We adopt the Solar motion of (U, V, W)$_\odot$ = (11.1,
12.24, 7.25) km~s$^{-1}$ \citep{Schonrich2010}, which is insensitive to the
metallicity gradient of the MW disc. The Galactic radius of the Sun is 8.2 kpc,
the solar circular speed is 232.8 km~s$^{-1}$, both suggested by
\citet{McMillan2017}.  Proper motions and distances of clusters are adopted
from \citet{Cantat-Gaudin2020b}.

\begin{figure*}[!htb]
    \centering
    \includegraphics[width=.98\textwidth]{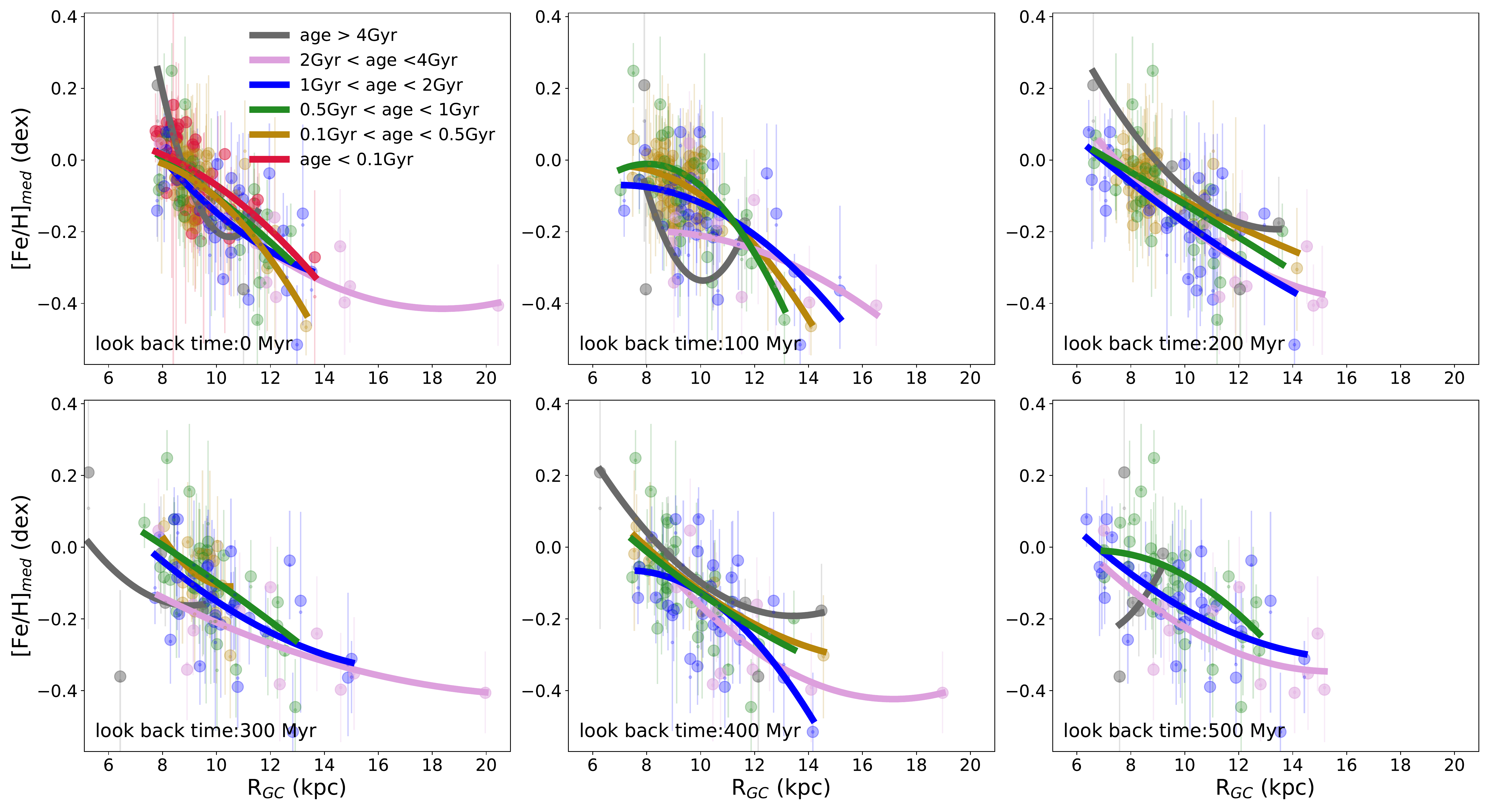}
    \caption{The clusters' median \feh as a function of the Galactocentric radius \rgc for all the LAMOST clusters with at least three {\sc flag=12} stars.
    Different panels show different look back times according to the orbit calculations.
    Clusters are divided into six groups according to age from the \citet{Cantat-Gaudin2020b} catalogue. 
    The solid curves are the second-order polynomial MCMC fittings of each age group.
   }
    \label{fig:feh_rgc_6fig}
\end{figure*}

Figure~\ref{fig:feh_rgc_6fig} shows the Galactic metallicity gradient traced by
LAMOST OCs in six snapshots during the past 500 Myr, that is,  from the present
time (look back time 0 Myr, uppermost left panel) back to the time of the last
passage of Sgr dSph. 
It is very difficult for LAMOST to observe stars toward
the Galactic centre, so almost all the LAMOST OCs are located with
Galactocentric radii $R_{\rm GC}\gtrapprox$8 kpc.  As discussed earlier,  only clusters with no less than three {\sc flag=12} stars are used in the following metallicity discussions.  Here we also
make a simplification that the cluster \feh does not change in its life time,
i.e., the microscopic diffusion in the stellar surface layer is not considered.
To investigate the metal enrichment in different epochs, we divide the LAMOST
OCs into six age groups (colour-coded).  A second-order polynomial fitting is
applied to each age group to measure their metallicity trend:

\begin{equation}\label{equ:fehrgc}
    \texttt{[Fe/H]} = {a \times R_{GC}^2 + b \times R_{GC} + c}
\end{equation}

For each fitting, we use a Markov chain Monte Carlo (MCMC)  method to consider
the \feh uncertainty of each cluster.  A number of 300 walkers and an
10,000-step MCMC are used in each fitting, based on the MCMC python package
\texttt{emcee} \citep{emcee}.  The parameters (a, b, c)  to  second-order polynomial fits
of the current time are listed in the \rgc column of Table
\ref{tab:mcmcpar}. 

In the literature, a linear fit has been widely adopted to describe the
metallicity radial gradient \citep[see e.g.][]{Friel2002, Bragaglia2006,
Carrera2019, Zhong2020,  Zhang2021}, but a break or a ``knee'' point is needed
at $\sim$ 12--14 kpc in order to properly take into account the outer disc
clusters \citep[see e.g.][]{Reddy2016, Donor2020, Spina2022}. Indeed, radial
metallicity distribution studies in other galaxies suggest that breaks and
changes of slopes are required in the linear fittings, and non-linear models
often fit better than linear models \citep{Scarano2013}. For this reason we use
the second-order polynomial model fittings to illustrated the metallicity
trends. From the six snapshots of the \feh trend in Fig.
\ref{fig:feh_rgc_6fig}, it is clear that, though generally speaking \feh is
higher at smaller \rgc, the metallicity gradients evolve among age groups,
instead of being a constant curve.  In other words, if we were observing OCs at
different times in the past 500 Myr, we would have seen very different pictures
of the metallicity gradient(s): It covers multiple shapes of being steep, flat,
with a turning point, and close to a straight line.

\begin{figure*}
    \centering
        \begin{subfigure}[]{0.48\textwidth}
        \includegraphics[width=\textwidth]{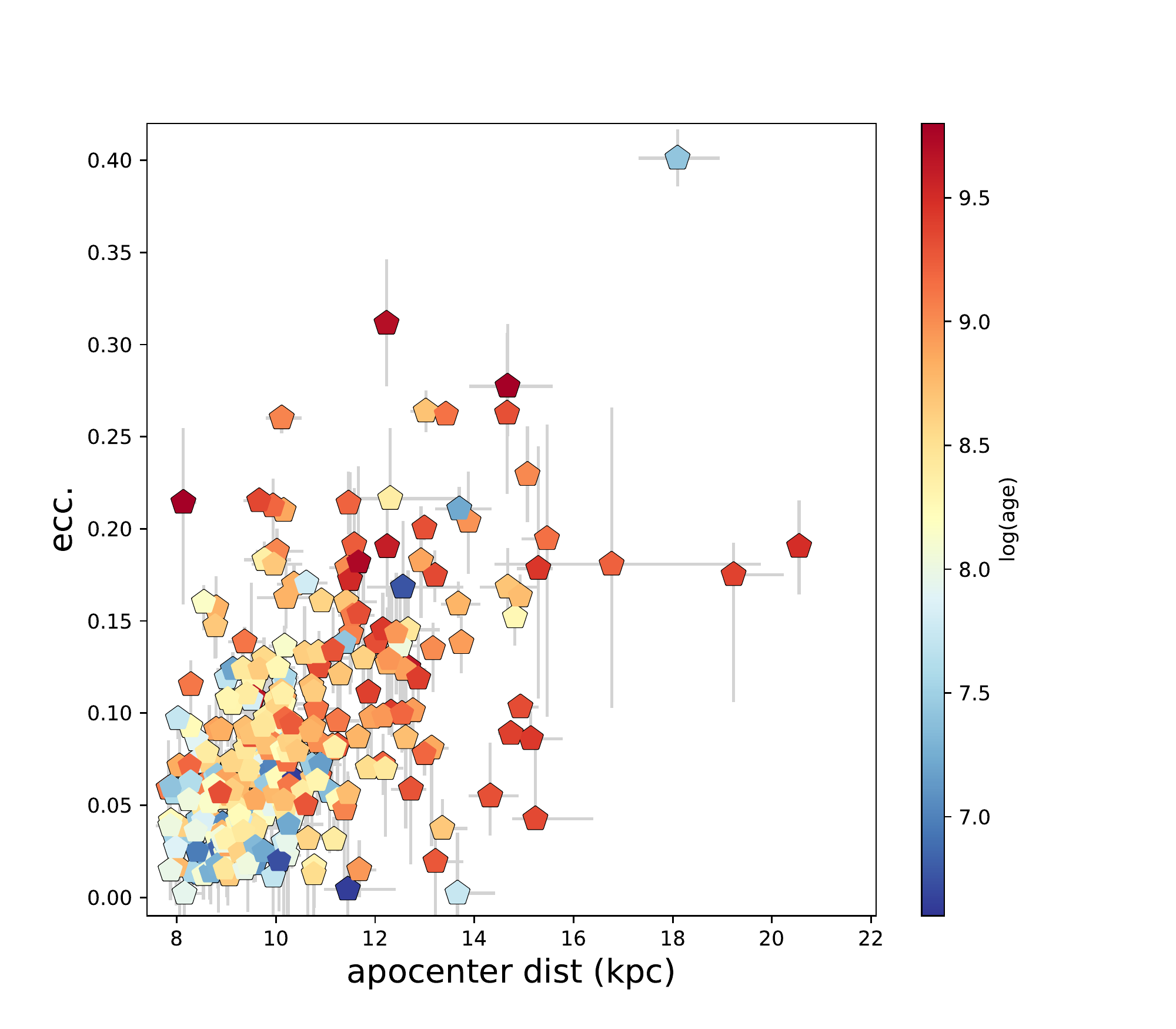}
    \end{subfigure}
    \begin{subfigure}[]{0.48\textwidth}
        \includegraphics[width=\textwidth]{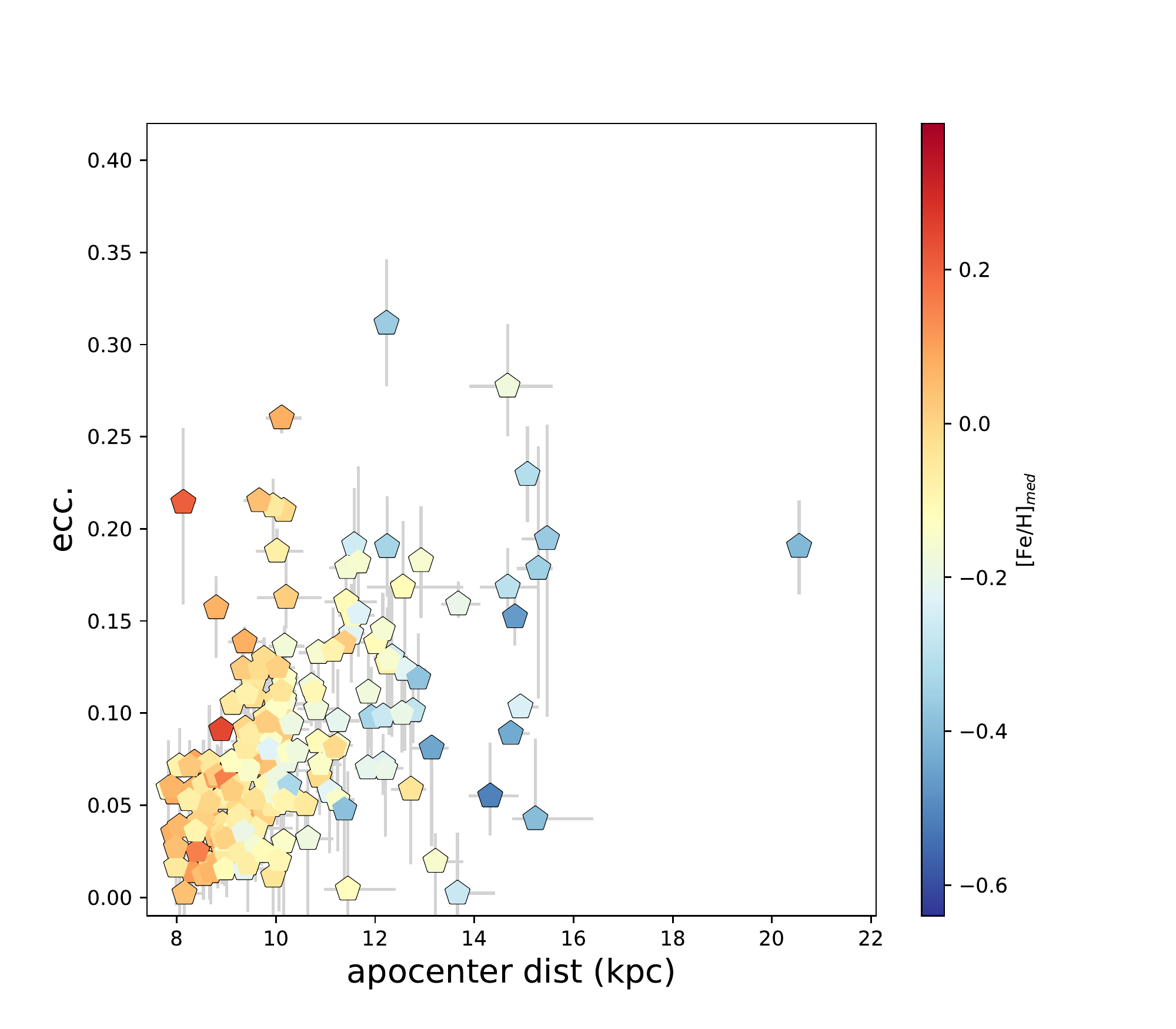}
    \end{subfigure}
    \caption{The clusters' eccentricity as a function of their Galactic apocentric distance.
    The colour code is the log value of the cluster age in the left panel,
    and cluster \feh in the right panel.
    Only clusters with multiple members are displayed in the left panel,
    while those with at least three {\sc flag=12} stars are displayed in the right panel.}
    \label{fig:ecc}
\end{figure*}

In order to interpret radial metallicity gradient variations for different age
groups, we modelled
the circular orbits of each cluster with \texttt{GALPOT}. Note that radial metallicity gradients are themselves trajectory projections  of cluster orbits. We find that the
clusters are not travelling in circular orbits. Figure~\ref{fig:ecc} shows the
eccentricity (ecc) of LAMOST OCs as a function of their Galactic apocentre
distance.
Though \texttt{GALPOT} does not give uncertainties of the orbit parameters,
we calculate the possible orbit parameter changes according to the scatter of clusters'  V${\sc _{rad}}$.

Most clusters have a non-circular orbit (ecc$>$0). For instance,
Berkeley\,29 has at present the highest \rgc= 20.44 kpc, being located on
the outermost disc in our catalog. However, its apocentric and pericentric
distances are 20.54 kpc and 13.92 kpc, respectively, with an eccentricity of
0.19.  The most eccentric cluster with at least three {\sc flag=12} member
stars in our sample is Berkeley\,32 (ecc = 0.31). It has a current \rgc of
11.00 kpc, but it travels to 12.23 kpc and 6.42 kpc as its apocentre and
pericentre, respectively. 

Generally speaking, the youngest clusters are more likely to travel with a more
circular orbit, while the oldest clusters have higher chances to have larger
eccentricity (see the left panel of Fig. \ref{fig:ecc}).  The metal-poorer open
clusters, whether or not they have a  circular orbit, can reach larger apocentric
distances than the metal-richer clusters (see the right panel of Fig.
\ref{fig:ecc}).  The non-circular orbit around the Galactic centre  results in
 clusters that constantly and periodically alter their \rgc between their
pericentre and apocentre.  The alteration period is shorter than one revolution
time  of the cluster, making the cluster travel back and forth several times in
one revolution around the Galactic centre \citep[see also discussions
in][]{Lepine2011}. All these facts make the present-day Galactocentric radius
\rgc not an invariant in the Galactic metallicity distribution investigations.

\begin{figure*}
    \centering
    \begin{subfigure}[]{0.49\textwidth}
    \includegraphics[width=\textwidth]{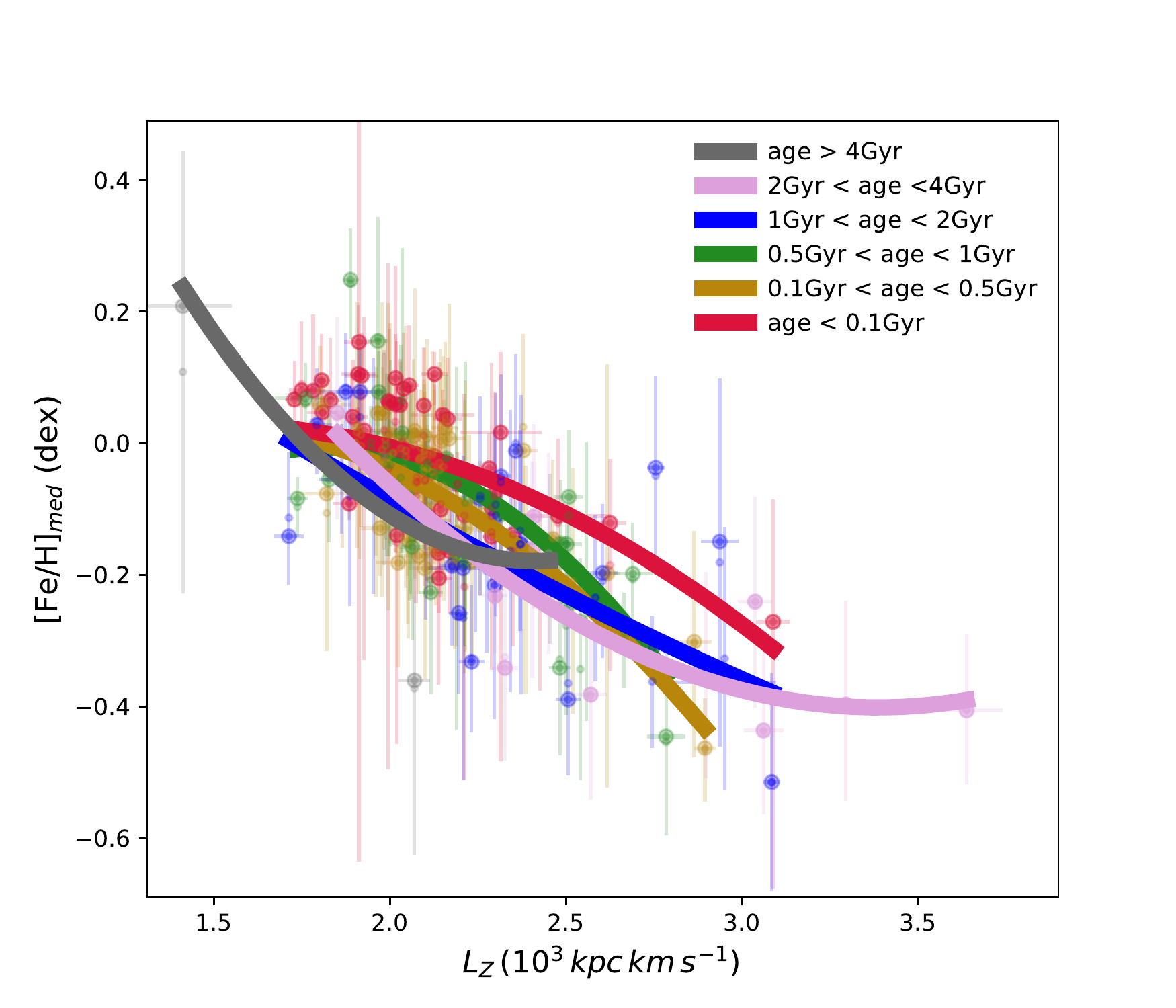}
    \end{subfigure}
    \begin{subfigure}[]{0.49\textwidth}
    \includegraphics[width=\textwidth]{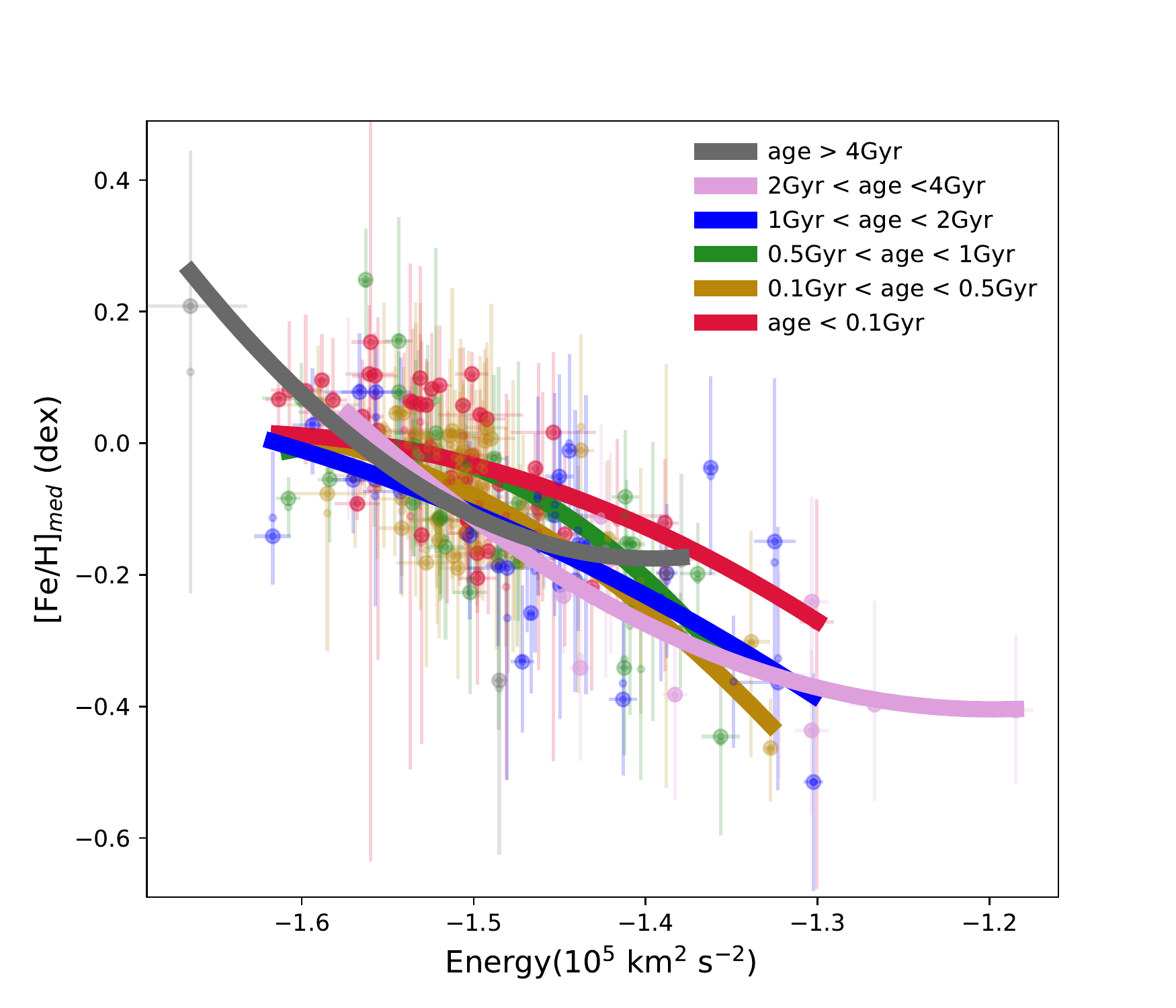}
    \end{subfigure}
    \caption{The clusters' median \feh as a function of their Z component of the Galactic angular momentum (left panel)
     and the total Galactic energy  (right panel) for all the LAMOST clusters with no less than three  {\sc flag=12} stars.
    The solid curves  are the second-order polynomial MCMC fittings of each age group. 
     }
    \label{fig:feh_lzen}
\end{figure*}

Thanks to the astrometry results provided by \gaia and radial velocity provided
by spectroscopic facilities such as LAMOST, we can now use more invariant
parameters such as the angular momenta and the dynamical energy to unfold the
whole story of the Galactic \feh evolution in the \gaia era.  In Fig.
\ref{fig:feh_lzen} we display the \feh distribution of the LAMOST OCs as a
function of the Z component of their angular momentum (left panel) and the
total dynamical energy (right panel). 
The corresponding uncertainties reflect the scatter of clusters'  V${\sc _{rad}}$.
These two parameters take into
account both the spatial positions and the motions, so they are constant for a cluster
in a given gravitational potential. We calculate them with \texttt{GALPOT}
following the same setups described before in the orbit integration part.  The
Z component of the angular momentum ($L_{\rm Z}$), i.e. the angular momentum
component that describes the motion around the Galactic centre on the Galactic
plane, is positive in the direction of the Galactic disc movement and has a
larger value at the outer disc. The dynamical energy (Energy) describes the
total energy of a cluster including kinetic and potential energy. For disc
stars with normal motions, a location further from the Galactic centre usually
leads to a higher value of the total energy.  

As shown in Fig. \ref{fig:feh_rgc_6fig}, OCs are divided into six groups based
on their current age and only clusters with at least three {\sc flag=12} stars
are included in the figure. The second-order polynomial MCMC fitting to each
age group is calculated with the same settings as described earlier. The
fitting parameters and their 16\% and 84\% distribution in the MCMC samples are
listed in Table. \ref{tab:mcmcpar}.  The metallicity distribution along $L_{\rm
Z}$ or energy could be considered as the Galactic metallicity gradient form in
the \gaia era.

From the \feh trend we could find that the very old age groups, those with age
$>$ 4 Gyr (grey curves) and within an age range of [2 Gyr, 4 Gyr] (plum
curves), both show a flat tail at their low metallicity end. Such flat metallicity 
tail has \feh$\sim-0.2$ dex for the age $>$ 4 Gyr group, starting from 
$L_{\rm Z}\sim$ 2.1$\times$10$^3$ kpc\,km\,s$^{-1}$ and energy 
$\sim-1.48\times10^5$ km$^2$\,s$^{-2}$,
and \feh$\sim-0.4~$ dex for the [2 Gyr, 4 Gyr] group, 
starting from $L_{\rm Z}\sim 3.\times10^3$ kpc\,km\,s$^{-1}$ and 
energy $\sim-1.3\times10^5$ km$^2$\,s$^{-2}$, respectively.  The other four
younger OC groups do not show such a flat tail feature.

The flat metallicity trend in the outer part of a galactic disc is known in the
literature, both in our own Galaxy \citep[see e.g.][]{Lepine2011,Andreuzzi2011,
Spina2022} and in other galaxies \citep{Scarano2013}. However, here we show
that the OC metallicity gradients appear to depend on their age, and the
two flat tails display two different metallicity ``plateaus''. 

The nature of the flat metallicity tail is still an open question.
\citet{Lepine2011} propose a hypothesis that clusters at the outer Galactic
disc with a flat \feh trend may have their birth place at a smaller \rgc. These
clusters travel between their orbit pericentre and apocentre, and are observed
in their present day position at a larger \rgc. This hypothesis could explain
the absence of young cluster at the outer Galactic disc and allow a simple
trend of metallicity along the Galactic radius. Our metallicity trend with
$L_{\rm Z}$ and energy, however, considers both spatial position and motions,
thus should not be affected by orbit.

The existence of the flat metallicity tail on the $L_{\rm Z}$ and energy plane,
especially the two different flat metallicity tails, must come from an
alternative mechanism. Another possible hypothesis proposed by
\citet{Lepine2011} is the gas flow from the relative inner region 
(the corotation radius) of the Galaxy
to the external regions, which brings gas with a slightly higher metallicity to the outer regions and flattens the metallicity gradients. 
The driven of the gas flow could be the gas interaction with the spiral potential perturbation \citep[see detail discussions and simulations in][]{lepine2001}.
With the current set of OC data, we are not able to confirm or exclude this hypothesis.
Detailed chemical abundances of the outer disc clusters may help to revive the
initial gas composition of these clusters. Whether the gas flow hypothesis can
also explain the other flat metallicity tail that of the age $>4$ Gyr clusters
requires further investigations.

Another key to understanding the metallicity evolution history of the flat metallicity tail is to investigate the origin of clusters in the Galactic outer disc, namely the clusters with high values of $L_{\rm
Z}$ and energy. Such clusters are visible on the right part of both panels in Fig.
\ref{fig:feh_lzen}. Are these clusters native residents of our
Galaxy, or do they have an extragalactic origin? For instance, the formation
and origin of Berkeley\,29 -- which is the outermost known disc cluster
with a \rgc of $\sim$20 kpc -- are discussed in the literature \citep{Carraro2009}. 
In their work, \citet{Carraro2009} find that the trailing tail of Sgr dSph
passes close to the location of Berkeley\,29, and their [Mg/Fe], [Ca/Fe]
abundances are similar to each other at the same \feh. They suggest that
Berkeley\,29 was formed in Sgr dSph and was left on the Galactic disc during
one of Sgr dSph's passages through our Galaxy about 5 Gyr ago. 
In fact, the mean metallicity of dwarf satellite galaxies may be higher than that of the original outskirts stars of the MW disc. 
After the satellite galaxy was disrupted by the MW tidal field, the stripped stars distribute mainly on
the outer region of MW due to their high kinetic energy and angular
momentum. 
These processes will increase the overall metallicity of the
outskirts and form a flat metallicity tail. 
In a follow-up paper (Chang in prep.), we will discuss in detail the role of minor mergers on forming the flat metallicity tail.

\begin{table}
\renewcommand\arraystretch{2}
\centering
\caption
{The second-order polynomial MCMC fitting parameters of the cluster \feh as a
function of the Galactic radius ($R_{\rm GC}$), the Z component of the
Galactic angular momentum ($L_{\rm Z}$), and the total Galactic energy (En)
for six age groups, respectively.
(a$_6$, b$_6$, c$_6$) stands for the age $>$ 4Gyr group,
(a$_5$, b$_5$, c$_5$) stands for the 2Gyr $<$ age $<$ 4Gyr group,
(a$_4$, b$_4$, c$_4$) stands for the 1Gyr $<$ age $<$ 2Gyr group,
(a$_3$, b$_3$, c$_3$) stands for the 0.5Gyr $<$ age $<$ 1Gyr group,
(a$_2$, b$_2$, c$_2$) stands for the 0.1Gyr $<$ age $<$ 0.5Gyr group,  and
(a$_1$, b$_1$, c$_1$) are the  age $<$ 0.1Gyr group.
The range of each parameter marks the 16\% and 84\% of the MCMC fitting sample distribution. 
}
\begin{tabular}{c|c|r|r|r}
\hline
\multicolumn{1}{c}{age (Gyr)}  &
\multicolumn{1}{c}{par.}  &
\multicolumn{1}{c}{R$_{GC}$ }  &
\multicolumn{1}{c}{ L$_Z / 10^3$ }  &
\multicolumn{1}{c}{ En$ / 10^5$}  \\
\hline
             &    a$_6$ &    0.066 $^{ +0.102 } _{-0.101} $    &    0.423  $^{ +0.706} _{ -0.763} $    &   6.343  $^{ +9.242} _{ -9.366 } $     \\
 $>4$        &    b$_6$ &    -1.379$^{ +2.021 } _{-2.038} $    &    -2.037 $^{ +3.133} _{ -2.890} $    &   17.821 $^{ +27.608} _{ -27.947} $     \\
             &    c$_6$ &    7.019 $^{ +9.951 } _{-9.878} $    &    2.278  $^{ +2.965} _{ -3.209} $    &   12.331 $^{ +20.589} _{ -20.830} $     \\
\hline                                                              
             &    a$_5$ &    0.004 $^{ +0.003 } _{-0.003} $    &    0.177  $^{ +0.153} _{ -0.143} $    &   3.203  $^{ +3.122} _{ -3.373 } $     \\
$[2,4]$      &    b$_5$ &    -0.143$^{ +0.090 } _{-0.090} $    &    -1.196 $^{ +0.809} _{ -0.866} $    &   7.673  $^{ +8.505} _{ -9.148 } $     \\
             &    c$_5$ &    0.894 $^{ +0.633 } _{-0.629} $    &    1.621  $^{ +1.191} _{ -1.126} $    &   4.198  $^{ +5.765} _{ -6.078 } $     \\
\hline                                                              
             &    a$_4$ &    0.005 $^{ +0.008 } _{-0.008} $    &    0.037  $^{ +0.167} _{ -0.229} $    &   -1.522 $^{ +3.596} _{ -3.235 } $     \\
$[1,2]$      &    b$_4$ &    -0.170$^{ +0.176 } _{-0.176} $    &    -0.459 $^{ +1.053} _{ -0.765} $    &   -5.670 $^{ +10.621} _{ -9.556 } $     \\
             &    c$_4$ &    1.026 $^{ +0.897 } _{-0.898} $    &    0.685  $^{ +0.872} _{ -1.181} $    &   -5.182 $^{ +7.819} _{ -7.043 } $     \\
\hline                                                              
             &    a$_3$ &    -0.002$^{ +0.009 } _{-0.009} $    &    -0.371 $^{ +0.187} _{ -0.197} $    &   -6.969 $^{ +3.223} _{ -3.243 } $     \\
$[0.5,1]$    &    b$_3$ &    -0.017$^{ +0.189 } _{-0.190} $    &    1.349  $^{ +0.869} _{ -0.825} $    &   -21.900$^{ +9.621} _{ -9.671 } $     \\
             &    c$_3$ &    0.274 $^{ +0.930 } _{-0.925} $    &    -1.233 $^{ +0.894} _{ -0.940} $    &   -17.206$^{ +7.164} _{ -7.196 } $     \\
\hline                                                              
             &    a$_2$ &    -0.009$^{ +0.008 } _{-0.008} $    &    -0.215 $^{ +0.235} _{ -0.173} $    &   -4.113 $^{ +2.934} _{ -3.074 } $     \\
$[0.1,0.5]$  &    b$_2$ &    0.122 $^{ +0.162 } _{-0.163} $    &    0.611  $^{ +0.811} _{ -1.095} $    &   -13.651$^{ +8.598} _{ -9.005 } $     \\
             &    c$_2$ &    -0.383$^{ +0.833 } _{-0.830} $    &    -0.402 $^{ +1.238} _{ -0.937} $    &   -11.297$^{ +6.292} _{ -6.590 } $     \\
\hline                                                              
             &    a$_1$ &    -0.005$^{ +0.012 } _{-0.011} $    &    -0.135 $^{ +0.252} _{ -0.210} $    &   -2.439 $^{ +4.097} _{ -3.798 } $     \\
$<0.1$       &    b$_1$ &    0.044 $^{ +0.205 } _{-0.223} $    &    0.406  $^{ +0.873} _{ -1.053} $    &   -8.025 $^{ +12.480} _{ -11.586} $     \\
             &    c$_1$ &    -0.027$^{ +1.032 } _{-0.951} $    &    -0.277 $^{ +1.097} _{ -0.904} $    &   -6.577 $^{ +9.508} _{ -8.830 } $     \\
\hline
\end{tabular}
\label{tab:mcmcpar}
\end{table}

One of the most popular and efficient methods to identify different Galactic
structures is the Integral Of Motion (IOM) space \citep[see discussions
in ][and references therein]{Helmi2020}. The main idea of this method is using
the total energy and the Z component of angular momentum as a long-term record.
Stars or clusters with a similar origin tend to cluster in the energy-$L_{\rm Z}$
space. 

\begin{figure}
    \centering
    \includegraphics[width=0.55
    \textwidth]{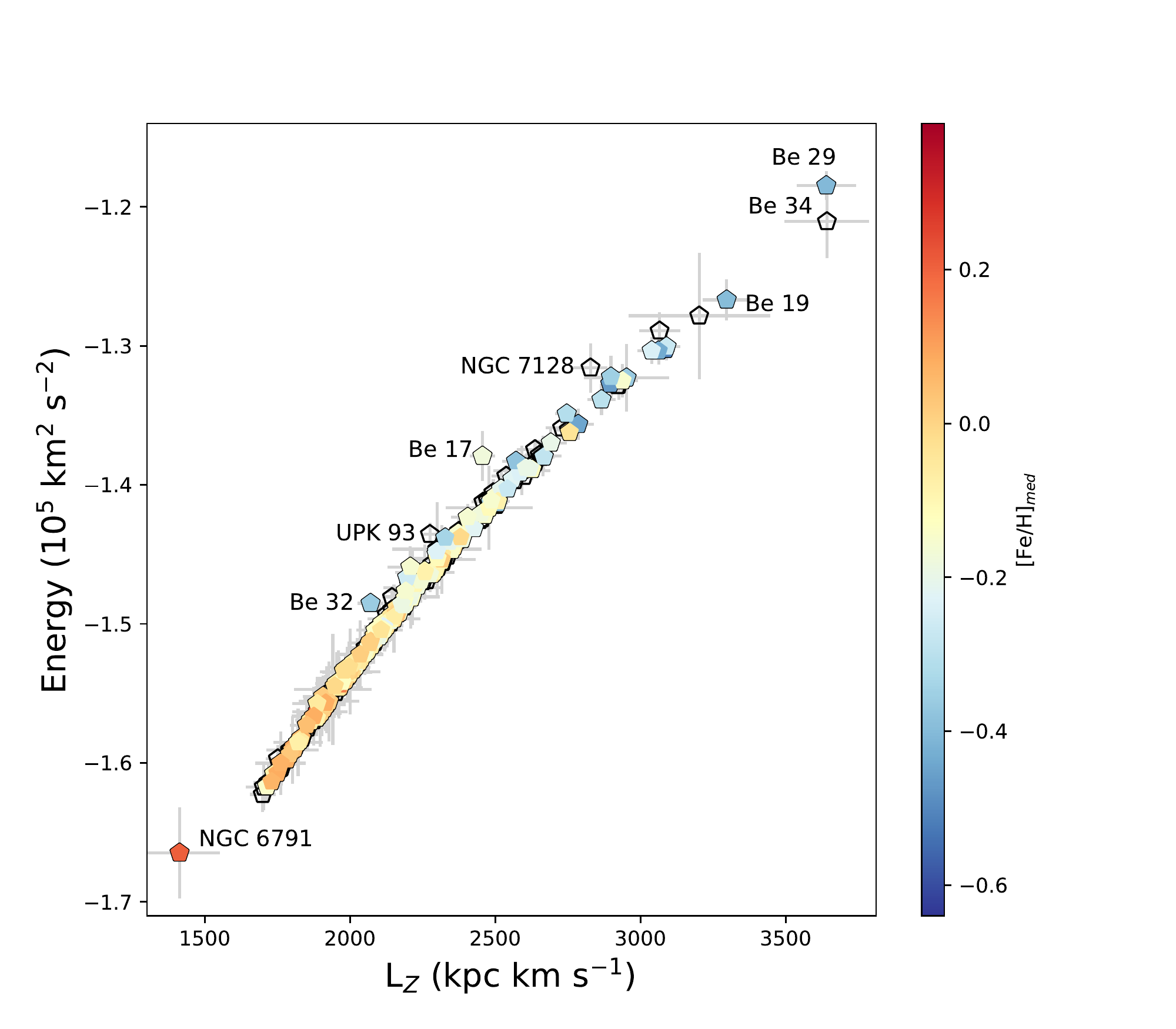}
    \caption{The Integral Of Motion (IOM) of LAMOST OCs with multiple members (N$_{\sc flag=1} \ge$ 2). 
           The \feh colour-code
            is applied to clusters with at least three {\sc flag=12} members.  The empty
            symbols are clusters with less than three {\sc flag=12} members.}
    \label{fig:lzen}
\end{figure}

Fig. \ref{fig:lzen} shows the IOM of LAMOST OCs.
Most clusters lie on a compact
region on the IOM space, which represents the dynamical property of the
Galactic thin disc. However, a few outliers do exist, in which we highlight a
cluster with a small value of both $L_{\rm Z}$ and Energy in the inner part of the
Galaxy (NGC\,6791), clusters with large values of $L_{\rm Z}$ and Energy in the
outer part of the Galaxy (Berkeley\,29, Berkeley\,34, and Berkeley\,19), and
clusters showing a departure from the main thin disc property (Berkeley\,32,
UPK\,39, Berkeley\,17, and NGC\,7128). These clusters must either have a
different origin, or have experienced different evolution compared to the
majority of normal OCs of the Galactic thin disc.

Similar to the Galactic field stars, clusters inhabit and travel on both sides
of the Galactic plane. The scale height of the Galactic disc is shaped by the
disc stars and OCs. It increases quickly from the Solar \rgc to the outer disc
on both sides of the plane. This increase in Z distance departure from the
Galactic plane  is referred as the disc flare. 
The flare has already been traced with different types of stars,
such as blue stragglers \citep[see e.g.][]{Thomas2019}, red clump stars
\citep[see e.g.][]{wan2017}, and red giants \citep[see e.g.][]{Wang2018}. 

\begin{figure}
    \centering
    \includegraphics[width=.48\textwidth]{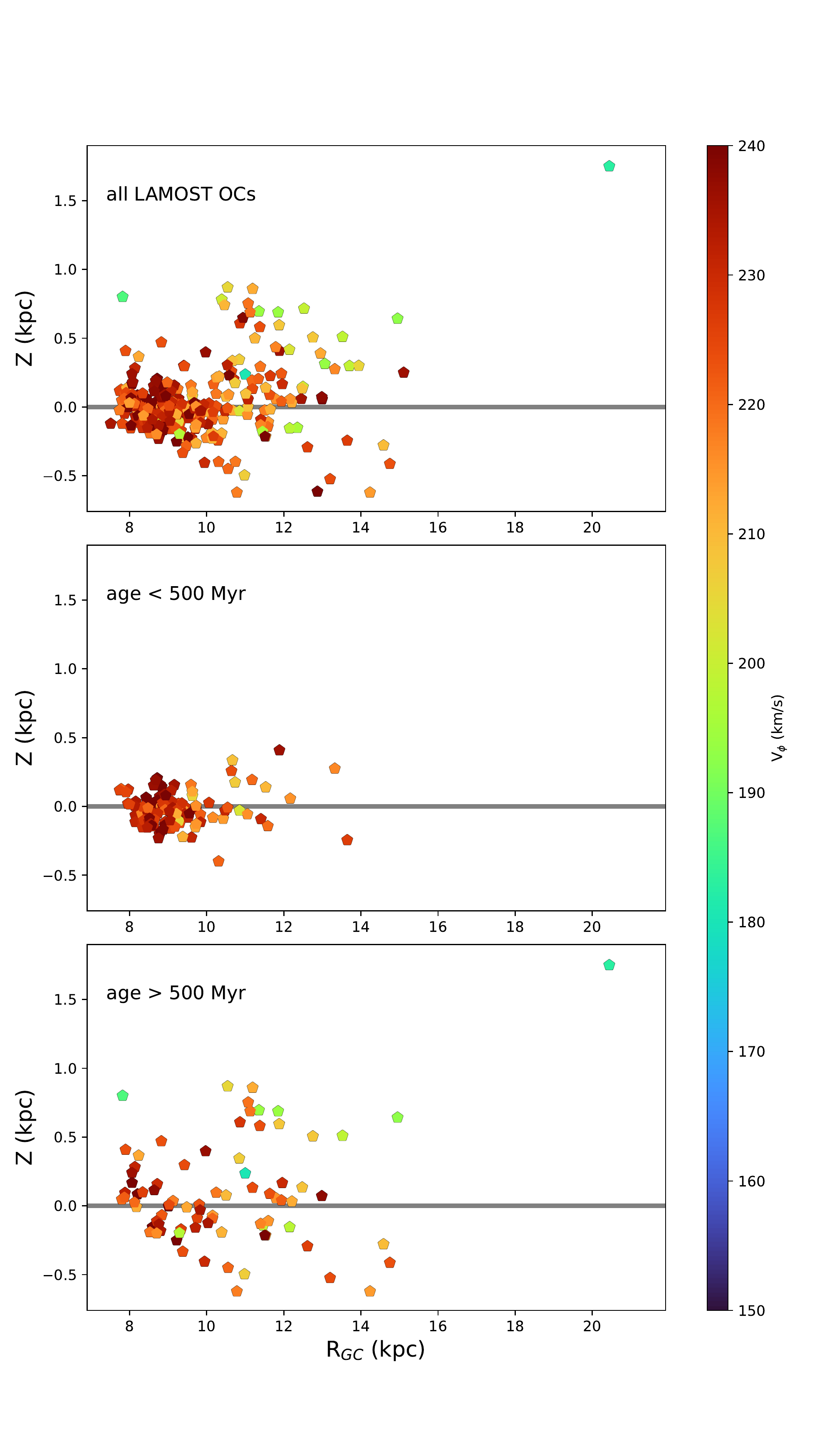}
    \caption{The present-day location of LAMOST OCs on the Z$_{GC}$-\rgc plane.
    The upper panel shows all LAMOST OCs, while the middle and lower panels show clusters born after the last passage of the Sgr dSph (age<500 Myr), and clusters born before the passage (age >500 Myr), respectively.
    Clusters are colour-coded with their azimuthal velocity V$_{\phi}$.}
    \label{fig:flare}
\end{figure}

Using simulations and LAMOST K giant stars, \citet{Xu2020} suggest that the
last impact of Sgr dSph contributes to the flare, and identify three different
branches of the flare with rotation velocity $V_\phi$.  According to their
results, the boundary of the flare is constructed by stars with a similar
$V_\phi$ to the main part of the disc stars.  The $V_\phi$ value for the flare boundary stars
is very different compared to halo stars at the same \rgc and stars on
the disc plane at larger \rgc.  In Fig.  \ref{fig:flare} we follow the analysis
of \citet{Xu2020} to investigate the LAMOST OCs distribution on the Z-\rgc
plane.  All clusters are colour-coded with their Galactic  V$_\phi$.

The upper panel shows all LAMOST OCs in our catalogue.
It is apparent that the clusters' Z distances increase toward the outer disc, indicating that OCs can also probe the disc flare.
The flare clusters with similar V{\rm $_\phi$} to clusters close to the Galactic plane show an asymmetric structure from the north side to the south side of the Galactic plane.
Two sequences of clusters are apparent below the Galactic plane,  similar to the south branch and main branch  identified by \citet{Xu2020}, but less extended.
\citet{Xu2020} simulate the disc evolution after the impact of Sgr dSph $\sim$500 Myr ago and conclude that the interaction between the Galactic disc and the dwarf galaxy contribute the flare.
To check if OCs are also under the influence of the Sgr dSph passage, we divided our LAMOST OCs into two categories: 
a young group with  age $< 500$ Myr, representing clusters born after the last passage of  Sgr dSph;
and an old group with age $>500$ Myr, representing clusters born before the passage.
The two categories are shown in the middle and lower panel of Fig. \ref{fig:flare}, respectively.
We expect clusters in the older category to travel in the Galaxy more like a solid body and behave similarly to field stars,
while clusters in the young category should be less affected by the interaction directly because they were gas clouds or even stars of a previous generation, when the impact took place.  
Indeed, clusters with age > 500 Myr shape the  OC flare boundary (see the lower panel of Fig. \ref{fig:flare}),
while clusters with age < 500 Myr are more concentrated along the Galactic plane with a small V$_\phi$ variation. 

We do notice that according to radial migration simulations \citep[see e.g.][]{Minchev2012}, secular evolution of the Galactic disc can increase the scale height in the outer disc and decrease the scale height in the inner region. Thus, it can also produce a flare structure. However, 
clusters move outward by gaining angular momentum and azimuthal velocity V$_\phi$ \citep{schonrich2009} while
the low V$_\phi$ of the LAMOST OCs in the flare region are difficult to reconcile with radial migration.
Considering their asymmetric distribution  above and under the Galactic plane, we suggest that the disc perturbation introduced by the last impact of Sgr dSph contributes to the OC flare.

\subsection{Connection with nearby molecular clouds}
\label{sec:clouds}

Young clusters are a link between the interstellar medium (ISM) and stellar evolution.
Since most stars are born in stellar clusters, linking young clusters to their surrounding molecular clouds offers a great help to understand star formation on a Galactic scale.

\begin{figure*}[!htb]
    \centering
    \begin{subfigure}[]{0.48\textwidth}
        \includegraphics[width=\textwidth]{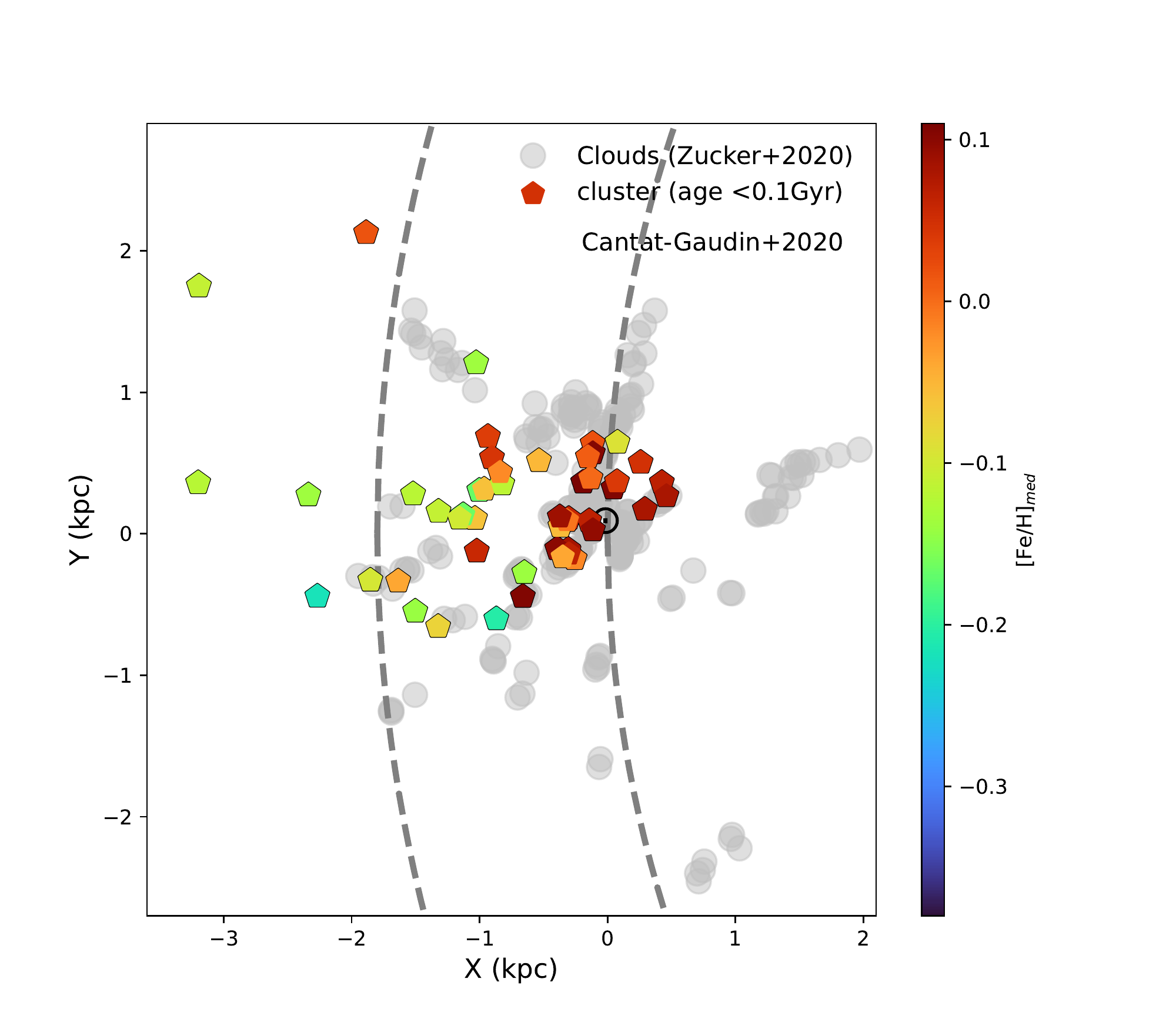}
    \end{subfigure}
    \begin{subfigure}[]{0.48\textwidth}
        \includegraphics[width=\textwidth]{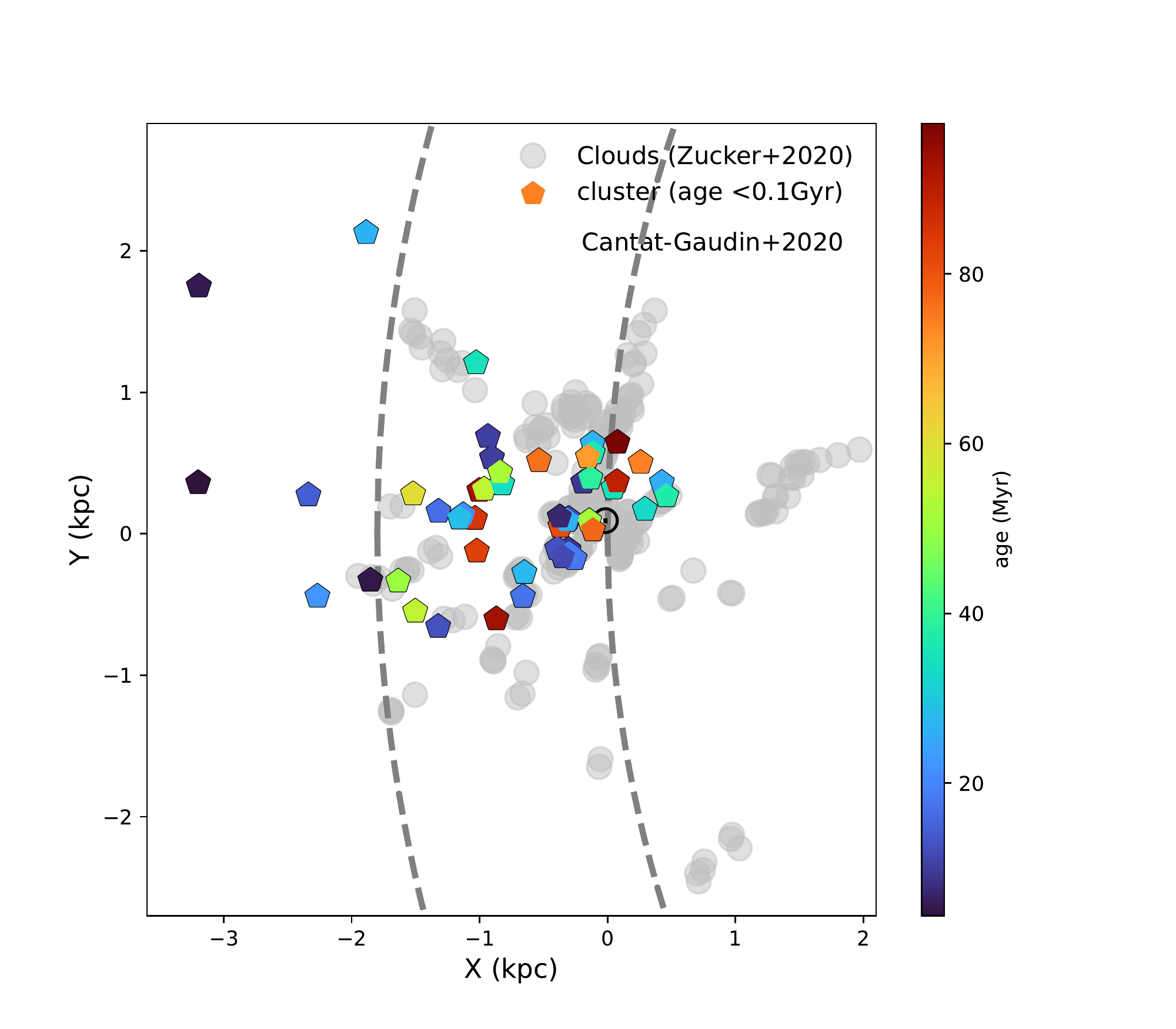}
    \end{subfigure}
    \caption{Positions of the age < 0.1 Gyr  clusters on the Galactic plane, 
    compared to the nearby molecular clouds from \citet{Zucker2020, clouddist}. 
    The two dashed circles indicate a Galactic radius of 8.2 kpc and 10 kpc.
    The clusters are colour-coded with their median \feh value in the left panel and with age in the right panel.
    Only clusters with more than three  {\sc flag=12} stars are displayed.
    }
    \label{fig:clouds}
\end{figure*}

Using \gaia DR2 distances and stellar optical and near-infrared photometry, \citet{Zucker2019} present a technique to determine distances of molecular clouds. 
\citet{Zucker2020} apply this method to $\sim$60 star-forming regions and molecular clouds within 2.5 kpc.
To investigate the connections between these clouds and young clusters,
 we show in Fig.\ref{fig:clouds} the positions of the \citet{Zucker2020} clouds in the Cartesian coordinate X-Y plane (centred on the Sun) and over-plot the young clusters  (age $<100$ Myr) of the LAMOST OC sample.
Many LAMOST young clusters show an overlap with the molecular clouds on the Galactic plane.
With such a young age and a similar Galactic position, they are very likely associated with each other.
Since it is easier to obtain metallicity from stars than from molecular clouds, a young cluster is a good probe to study the metallicity evolution of the surrounding molecular clouds.
We colour-code clusters in Fig.\ref{fig:clouds}  with \feh in the left panel and cluster age in the right panel.
It is clear that even for clusters that share a similar position on the Galactic plane, their \feh and age differ.
This indicates that there are multiple star formation epochs during the past 100 Myr in the Solar neighbourhood, with gas metallicity covering a $\Delta$\feh$\sim$0.4 dex (i.e. about 2.7 times difference in metallicity).
Whether this difference is due to inhomogeneous mixing in the giant molecular cloud, or because of fast star formation and pollution in the past 100 Myr, requires further investigations.

\begin{figure}[!htb]
    \centering
    \includegraphics[width=.48\textwidth]{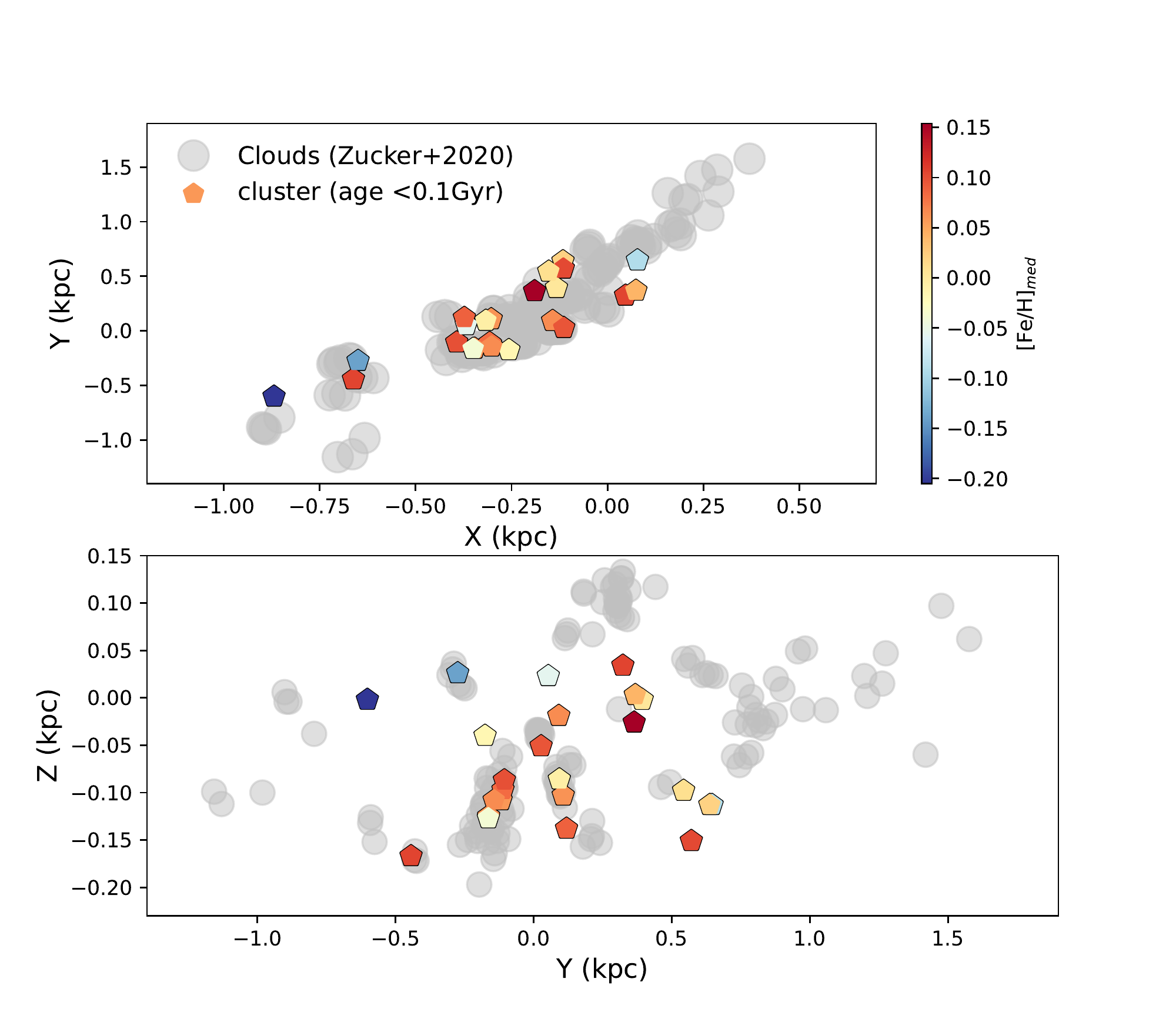}
    \caption{\feh distribution of clusters along the Radcliffe gas wave.
    The upper panel is a bird's eye view and the lower panel is a side view along the Cartesian Y axis.
    Clusters are colour-coded with their \feh values.}
    \label{fig:wave}
\end{figure}

To study the metallicity distribution along a sequence of nearby molecular clouds,
we investigate OCs around the Radcliffe  wave,  which is a coherent gaseous wave-like structure in the solar neighbourhood reported by \citet{Alves2020} based on the molecular cloud distance catalogue  \citep{Zucker2019, Zucker2020}.

Following \citet{Alves2020}, we select 24 LAMOST young OCs along the Radcliffe  wave. 
Table \ref{tab:waveoc} lists the Galactic coordinates, Cartesian coordinate XYZ, age, and \feh of these clusters.
They are located along the gas wave clouds in three-dimensional space.
Figure~\ref{fig:wave} shows the spatial distribution of these clusters together with the Radcliffe  wave clouds.
In both the Cartesian X-Y (upper panel) and Y-Z (lower panel) frame, these clusters show a high spatial consistency with the wave clouds.
The metallicities of these clusters, with the higher values in the more central part of the wave, cover a non-negligible range, from sub-solar to super-solar.
The highest \feh is about 1.4 times  the Solar metallicity, and the lowest \feh is about 0.6 times the Solar metallicity.
The \feh distribution of these clusters could be considered as the metallicity distribution of the Radcliffe  wave in the past star formation epochs.

\begin{table*}
\centering
\setlength{\tabcolsep}{1mm}
\caption{The Galactic coordinates, age, Cartesian coordinate XYZ (centred on the Sun), age, and \feh of the 24 very young LAMOST OCs along the Radcliffe  wave.}
\begin{tabular}{lrrrrrrr}
\hline
\multicolumn{1}{c}{cluster}  &
\multicolumn{1}{c}{Glon ($\degree$)}  &
\multicolumn{1}{c}{Glat ($\degree$)}  &
\multicolumn{1}{c}{age (Myr)}  &
\multicolumn{1}{c}{X (pc)}  &
\multicolumn{1}{c}{Y (pc)}  &
\multicolumn{1}{c}{Z (pc)}  &
\multicolumn{1}{c}{[Fe/H]}  \\
\hline
 ASCC\_16      & 201.139 & -18.37 & 13.48 & -305 & -118 & -108 & 0.06  $\pm$ 0.07  \\
 ASCC\_21      & 199.938 & -16.60 & 8.91  & -307 & -111 & -97  & 0.08  $\pm$ 0.10  \\
 ASCC\_29      & 214.743 & -0.13  & 93.32 & -869 & -602 & -2   & -0.21 $\pm$ 0.06  \\
 Alessi\_20    & 117.615 & -3.70  & 9.33  & -190 & 364  & -26  & 0.15  $\pm$ 0.57  \\
 Collinder\_69 & 195.162 & -12.05 & 12.58 & -393 & -106 & -87  & 0.10  $\pm$ 0.22  \\
 Gulliver\_6   & 205.246 & -18.14 & 16.59 & -346 & -163 & -125 & 0.06  $\pm$ 0.11  \\
 IC\_348       & 160.486 & -17.81 & 11.74 & -303 & 107  & -103 & 0.06  $\pm$ 0.13  \\
 Melotte\_20   & 147.357 & -6.40  & 51.28 & -143 & 91   & -19  & 0.07  $\pm$ 0.09  \\
 Melotte\_22   & 166.462 & -23.61 & 77.62 & -114 & 27   & -51  & 0.10  $\pm$ 0.07  \\
 NGC\_2183     & 213.898 & -11.84 & 17.37 & -662 & -444 & -167 & 0.11  $\pm$ 0.04  \\
 NGC\_2232     & 214.458 & -7.47  & 17.78 & -257 & -176 & -40  & -0.02 $\pm$ 0.12  \\
 NGC\_2264     & 202.941 & 2.17   & 27.54 & -650 & -275 & 26   & -0.14 $\pm$ 0.31  \\
 NGC\_7063     & 83.0930 & -9.89  & 97.72 & 78   & 646  & -113 & -0.09 $\pm$ 0.03  \\
 RSG\_5        & 81.7190 & 6.10   & 34.67 & 47   & 323  & 34   & 0.11  $\pm$ 0.11  \\
 RSG\_7        & 108.781 & -0.32  & 38.90 & -133 & 393  & -2   & 0.00  $\pm$ 0.09  \\
 Roslund\_6    & 78.4950 & 0.58   & 89.12 & 74   & 368  & 3    & 0.04  $\pm$ 0.10  \\
 Stock\_10     & 171.714 & 3.56   & 81.28 & -368 & 53   & 23   & -0.06 $\pm$ 0.10  \\
 UPK\_166      & 100.382 & -9.91  & 26.91 & -116 & 638  & -113 & 0.02  $\pm$ 0.26  \\
 UPK\_168      & 101.455 & -14.58 & 36.30 & -115 & 571  & -151 & 0.10  $\pm$ 0.10  \\
 UPK\_185      & 105.807 & -9.94  & 70.79 & -153 & 543  & -98  & 0.01  $\pm$ 0.04  \\
 UBC\_17a      & 205.335 & -18.02 & 18.62 & -302 & -143 & -108 & 0.06  $\pm$ 0.38  \\
 UBC\_31       & 163.527 & -14.72 & 26.30 & -317 & 93   & -86  & -0.01 $\pm$ 0.08  \\
 UBC\_17b      & 205.142 & -18.18 & 11.48 & -350 & -164 & -127 & -0.04 $\pm$ 0.09  \\
 UBC\_19       & 162.215 & -19.48 & 6.91  & -373 & 119  & -138 & 0.09  $\pm$ 0.10  \\
\hline
\end{tabular}
\label{tab:waveoc}
\end{table*}

\section{Summary}
\label{sec:sum}

Using high quality open cluster membership based on \gaia data  \citep{Cantat-Gaudin2020b} and spectroscopic results from LAMOST DR8, we obtain a LAMOST OC catalogue with 386 open clusters. 
The radial velocity and metallicity of these clusters are determined homogeneously. 
To our knowledge, this is the first radial velocity determination for 44 clusters and the first spectroscopic \feh determination for 137 clusters.
Among the clusters with newly-obtained \feh, 63 ones are based on at least three member stars with high quality stellar parameter determinations. 

The cluster parameter determinations are based on Monte Carlo sampling method.
Member stars used for cluster V${\sc _{rad}}$ determinations are marked with flag=1,
those used for both  V${\sc _{rad}}$ and \feh determinations are marked with  {\sc flag=12}.
The final LAMOST OC parameter catalogue, together with a table of member stars with quality control flag and LAMOST stellar parameters, is available on CDS.

During the quality control process to select {\sc flag=12} member stars, we notice a systematic issue of LAMOST cool main sequence stars.
Both the surface gravity \logg and iron abundance \feh are under-estimated for main sequence stars with \teff $\lessapprox$ 5000 K.
The problem appears in all clusters that have member stars in this range.
These stars are not considered in the cluster parameter determinations.
We suggest to take this problem into account also in studies using LAMOST field stars.

Using LAMOST OCs as tracers, we further study the Galactic metallicity distribution and the Galactic disc dynamical properties.
We calculate the orbit of the clusters and trace their Galactic metallicity gradient evolution in the past 500 Myr.
Since most of the clusters have a non-circular orbit around the Galactic centre, we suggest to use the Z component of the angular momentum L$_Z$ or the total dynamical energy instead of the current Galactic radius \rgc to describe the Galactic metallicity gradient. 
With these two forms of metallicity gradient, we find two  flat metallicity trend tails for very old OCs $>$ 4 Gyr and for OCs in the [2 Gyr, 4 Gyr] age range.

We also investigate the OC metallicity distribution in the IOM space.
Most of the LAMOST OCs lie in a tight line following the Galactic disc dynamics, while some outlier clusters show different \feh compared to other clusters in similar IOM space. 

LAMOST OCs can  be used to trace the Galactic disc flare.
The OC flare shows an asymmetric structure above and under the Galactic plane,  similar to the field star flare in the literature using LAMOST K giants \citep{Xu2020}.
Based on the morphology and V$_\phi$ distribution of the flare, we suggest that the last impact of Sgr dSph contribute to the OC flare. 

The very young LAMOST OCs (age < 100 Myr) are compared to nearby molecular clouds.
These young clusters, which are spatially overlapping  molecular clouds, can be used to probe star formation history and metallicity evolution of the molecular clouds.
We find that even for young clusters with very similar Galactic position, 
their \feh and age could differ by a factor of three. 
This indicates possible inhomogeneous mixing in local ISM \citep[see similar discussions in ][]{DeCia2021} or multiple star formations in OCs during short time scales.
Using 24 very young clusters along the Radcliffe wave, we present the metallicity distributions of the gas wave.
The nature of this metallicity distribution and its connection to the formation of the Radcliffe wave require further investigations.

\begin{acknowledgements}

X.F. acknowledges the support of China Postdoctoral Science Foundation No. 2020M670023. 
X.F. and A.B. acknowledge funding from the Italian MIUR through PREMIALE 2016 MITiC.
X.F. and H.Z. thanks the support of the National Key R\&D Program of China No. 2019YFA0405500 and the National Natural Science Foundation of China (NSFC) under grant No.11973001, 12090040, and 12090044. 
X.F., Z.Y.Z, C.L, Y.C, and J.Z  acknowledge the science research grants from the China Manned Space Project with NO.CMS-CSST-2021-A08.
J.Z. thanks the support of NSFC No. 12073060.
Y.C. acknowledges the support of NSFC under grant No. 12003001.
L.L. thanks the support of the UCAS Joint PHD Training Program.
L.C. and J.Z. acknowledges the support of NSFC through grants 12090040and12090042.
Z.Y.Z acknowledges the support of NSFC under grants No. 12041305, 12173016, and the Program for Innovative Talents, Entrepreneur in Jiangsu.

This work benefited from the International Space Science Institute (ISSI/ISSI-BJ) in Bern and Beijing, thanks to the funding of the team “Chemical abundances in the ISM: the litmus test of stellar IMF variations in galaxies across cosmic time” (PI D. Romano and Z-Y. Zhang). 
Guoshoujing Telescope (the Large Sky Area Multi-Object Fiber Spectroscopic Telescope LAMOST) is a National Major Scientific Project built by the Chinese Academy of Sciences. Funding for the project has been provided by the National Development and Reform Commission. LAMOST is operated and managed by the National Astronomical Observatories, Chinese Academy of Sciences. 
This work has made use of data from the European Space Agency (ESA) mission {\it Gaia} (\url{https://www.cosmos.esa.int/gaia}), processed by the {\it Gaia} Data Processing and Analysis Consortium (DPAC,
\url{https://www.cosmos.esa.int/web/gaia/dpac/consortium}). 
Funding for the DPAC has been provided by national institutions, in particular the institutions participating in the {\it Gaia} Multilateral Agreement. This research has made use of the TOPCAT catalogue handling and plotting tool \citep{topcat,topcat2017}; of the Simbad database and the VizieR catalogue access tool, CDS, Strasbourg, France \citep{Ochsenbein2000}; of NASA's Astrophysics Data System,
and is supported by High-performance Computing Platform of Peking University.
\end{acknowledgements}

   \bibliographystyle{aa} 
   \bibliography{lamostoc.bib} 

\end{document}